\documentclass[showpacs,twocolumn,preprintnumbers,amsmath,amssymb]{revtex4}

\usepackage{graphicx}
\usepackage{dcolumn}
\usepackage{bm}
\usepackage{ulem}

\newcommand{\bq}   {\begin{equation}}
\newcommand{\eq}   {\end{equation}}
\newcommand{\bqa}  {\begin{eqnarray}}
\newcommand{\eqa}  {\end{eqnarray}}
\newcommand{\nn}   {\nonumber \\}

\def\be         {\begin{equation}}
\def\ee         {\end{equation}}
\def\bea        {\begin{eqnarray}}
\def\eea        {\end{eqnarray}}
\def\bnn        {\begin{eqnarray*}}
\def\enn        {\end{eqnarray*}}

\def\tr         {\mathrm{tr}}
\begin{document}

\title{Spin-liquid Mott quantum criticality in two dimensions: Destabilization of a spinon Fermi surface and emergence of one-dimensional spin dynamics}
\author{Jae-Ho Han, Yong-Heum Cho, and Ki-Seok Kim}
\affiliation{Department of Physics, POSTECH, Pohang, Gyeongbuk 37673, Korea}

\date{\today}

\begin{abstract}
Resorting to a recently developed theoretical device called dimensional regularization for quantum criticality with a Fermi surface, we examine a metal-insulator quantum phase transition from a Landau's Fermi-liquid state to a U(1) spin-liquid phase with a spinon Fermi surface in two dimensions. Unfortunately, we fail to approach the spin-liquid Mott quantum critical point from the U(1) spin-liquid state within the dimensional regularization technique. Self-interactions between charge fluctuations called holons are not screened, which shows a run-away renormalization group flow, interpreted as holons remain gapped. This leads us to consider another fixed point, where the spinon Fermi surface can be destabilized across the Mott transition. Based on this conjecture, we reveal the nature of the spin-liquid Mott quantum critical point: Dimensional reduction to one dimension occurs for spin dynamics described by spinons. As a result, Landau damping for both spin and charge dynamics disappear in the vicinity of the Mott quantum critical point. When the flavor number of holons is over its critical value, an interacting fixed point appears to be identified with an inverted XY universality class, controlled within the dimensional regularization technique. On the other hand, a fluctuation-driven first order metal-insulator transition results when it is below the critical number. We propose that the destabilization of a spinon Fermi surface and the emergence of one-dimensional spin dynamics near the spin-liquid Mott quantum critical point can be checked out by spin susceptibility with a $2 k_{F}$ transfer momentum, where $k_{F}$ is a Fermi momentum in the U(1) spin-liquid state: The absence of Landau damping in U(1) gauge fluctuations gives rise to a divergent behavior at zero temperature while it vanishes in the presence of a spinon Fermi surface.
\end{abstract}


\maketitle

\section{Introduction}

Hertz-Moriya-Millis theory is a standard theoretical framework for quantum criticality in metals \cite{Hertz,Moriya,Millis}. Within the self-consistent random-phase-approximation (RPA) analysis, critical order-parameter fluctuations become overdamped, referred to as Landau damping. As a result, the dynamical critical exponent is enhanced to result in the fact that critical dynamics of order parameter fluctuations is essentially mean-field-like since the critical field theory is above the upper critical dimension. Such mean-field-type critical dynamics does not respect the hyperscaling relation due to the presence of a dangerously irrelevant operator \cite{HFQCP_Review}. This gives rise to the violation of the $\omega/T$ scaling behavior of the dynamical susceptibility for critical fluctuations, where $\omega$ is frequency and $T$ is temperature. Even if low-energy critical electrons are taken into account fully self-consistently in the RPA level, the mean-field-type scaling theory with a dangerously irrelevant operator remains essentially unchanged. On the other hand, the scaling dimension of the dangerously irrelevant operator becomes more positive and thus, more irrelevant in the presence of low-energy critical electrons.

The Yukawa coupling between low-energy electrons and critical order-parameter fluctuations is marginal at the critical point within the self-consistent RPA analysis. Then, such an approximation scheme can be dangerous in the case when the fixed-point value does not reside within the convergence area for the self-consistent RPA analysis. In order to justify this approximation scheme, one may increase spin degeneracy from $\sigma = \uparrow, \downarrow$ to $\sigma = 1, ..., N$. Then, the interaction vertex is reduced from $g$ to $g / \sqrt{N}$, and the self-consistent RPA analysis seems to be justified in the $N \rightarrow \infty$ limit. Any vertex corrections give rise to higher order contributions in $\mathcal{O}(1/N)$ and thus, self-energy corrections turn out to be in the leading order, referred to as the $1/N$ expansion \cite{Pepin_Chubukov}.

Recently, S.-S. Lee has shown that the self-consistent RPA analysis cannot be justified even in the $N \rightarrow \infty$ limit \cite{SSL_Breakdown_N}. He starts from the Hertz-Moriya-Millis fixed point in the self-consistent RPA framework: Critical boson dynamics is described by Landau damping with the dynamical critical exponent $z = 3$ and the dynamics of low-energy critical electrons is given by the following non-Fermi liquid self-energy correction, $\Sigma(i\omega) \sim i (g^{2}/N) \mbox{sgn}(\omega)|\omega|^{2/3}$. Performing the scaling analysis that makes the self-consistent RPA critical theory scale-invariant, he finds two essential aspects \cite{SSL_Breakdown_N_Before}: First, the angular part of the momentum integral acquires an anomalous scaling dimension. Second, such an anomalous scaling exponent justifies the double-patch construction as a minimal effective field theory. In particular, the overlapping region between two different patches is shown to vanish in the infrared (IR) limit \cite{Patch_Construction}. Based on this effective critical field theory, S.-S. Lee investigated the stability of the self-consistent RPA fixed point. It turns out that the presence of the $1/N$ factor in the non-Fermi liquid self-energy correction spoils the structure of the $1/N$ expansion \cite{SSL_Breakdown_N}. In particular, he suggests the double-line representation for Feynman diagrams, where boson fluctuations are given by double lines and fermion excitations are described by single lines. As a result, he reveals that the number of decoupled fermion loops correspond to the enhancement factor of $N$, originating from the $1/N$ factor of the self-energy correction and identified with Fermi surface fluctuations. Although this counting rule turns out to break down beyond the single-patch approximation \cite{Max_Sachdev_Nematic,Max_Sachdev_SDW}, this study proposes that vertex corrections should be taken into account properly in order to describe critical dynamics of low-energy order-parameter fluctuations and fermion excitations. We do not know how to incorporate such vertex corrections into the quantum criticality of the Hertz-Moriya-Millis fixed point systematically based on the field theoretical approach.

In order to make the physical description of metallic quantum criticality mathematically controllable, theoreticians have tried to find another expansion parameter beyond the $N \rightarrow \infty$ limit \cite{Nonlocal_Interaction,SSL_Dimensional_Regularization_Nematic,SSL_Dimensional_Regularization_SDW}. Here, we focus on the dimensional regularization technique \cite{SSL_Dimensional_Regularization_Nematic,SSL_Dimensional_Regularization_SDW}. The dimensional regularization technique is well known for bosonic quantum criticality \cite{Dimensional_Regularization_Review}. Although its conceptual aspect is completely clear, a concrete manipulation has not been performed for the Fermi-surface problem. In references of \cite{SSL_Dimensional_Regularization_Nematic,SSL_Dimensional_Regularization_SDW}, S.-S. Lee proposed an interesting field-theoretical setup for the dimensional regularization technique to the Fermi-surface problem. We would like to call it ``graphenization" of the Fermi-surface problem. Maintaining the dimension of a Fermi surface with one dimension, he devises how to put the problem of spatially two-dimensional metallic quantum criticality into $d-$dimensions. Here, $d$ counts only the spatial part. Suppose Ising nematic quantum criticality in two dimensions. In order to maintain the shape of the one-dimensional Fermi surface in three dimensions for example, one should gap out the band structure of electrons along the $z-$dimension. The resulting band structure turns out to describe $p_{z}-$wave superconductivity in thee dimensions \cite{SSL_Dimensional_Regularization_Nematic}. Within the dimensional regularization technique, the nematic quantum criticality in two dimensions can be achieved from the band structure of $p_{z}-$wave superconductivity in three dimensions. In this situation the upper critical dimension of the Yukawa coupling between low-energy electrons and critical Ising nematic fluctuations is $d_{c} = 5/2$. Although it is questionable whether or not we are solving the same problem as the originally suggested one, the physical description is now completely justified at least mathematically, performing the renormalization group analysis in a slightly lower dimension than the upper critical dimension, i.e., $d = d_{c} - \varepsilon$ with $d_{c} = 5/2$ and $\varepsilon = 1/2$.

The ``graphenized" effective field theory of the double patch construction in $d-$dimensions allows an interacting fixed point for the Yukawa coupling constant. Critical dynamics of order-parameter fluctuations and non-Fermi liquid physics of low-energy electrons are described by the renormalization group analysis based on the dimensional regularization technique. Solving Callan-Symanzik equations gives scaling theories for correlation functions, identifying the nature of this novel interacting fixed point. The resulting interacting fixed point differs from the Hertz-Moriya-Millis critical point given by the self-consistent RPA analysis \cite{SSL_Dimensional_Regularization_Nematic,SSL_Dimensional_Regularization_SDW}. An essential point of the dimensional regularization technique is that the dynamical critical exponent is much less than the value of the self-consistent RPA theory. This is certainly expected due to the presence of pseudogap in the graphenization technique, responsible for the appearance of an interacting fixed point. However, it is not completely clear at all whether or not such an interacting fixed point reflects the nature of the originally proposed metallic quantum critical point, frankly speaking. Suppose the Kondo problem. It is well understood that the nature of the quantum critical point between the local moment phase and the local Fermi-liquid state in the pseudogap Kondo model \cite{Vojta_PKM} differs from that in a normal metallic host \cite{KE_Review}, where such a quantum critical point does not exist in the latter case. However, we reach the same renormalization group equation for the Kondo coupling constant if the pseudogap density-of-states parameter sets to vanish and recover the finite density of states as in normal metals.

Here, we adopt the dimensional regularization technique for the renormalization group analysis. In this study, we consider an insulator-metal transition from a U(1) spin-liquid state with a spinon Fermi surface to a Fermi-liquid phase, given by the Higgs transition of bosonic charge degrees of freedom referred to as holons \cite{U1SR}. The effective field theory for this spin-liquid Mott quantum criticality is as follows: First, critical spin dynamics is described by spin doublets interacting through low lying spin-singlet fluctuations, where spin doublets form a Fermi surface of spinons and low lying spin-singlets are expressed by U(1) gauge fluctuations. This Fermi-surface problem is exactly the same as that solved before by the dimensional regularization technique \cite{SSL_Dimensional_Regularization_Nematic}. Second, critical charge dynamics is described by sound modes interacting via low lying spin-singlet fluctuations, where sound modes are given by bosonic holons with the relativistic spectrum. This critical charge dynamics has never been taken into account on equal footing with the Fermi-surface problem in a controllable way.

We start from the U(1) spin-liquid interacting fixed point as intensively discussed above, which occurs from the dimensional regularization technique for the sector of the spinon-gauge field problem, essentially the same as the Ising nematic quantum criticality problem \cite{SSL_Dimensional_Regularization_Nematic}. The appearance of such an interacting fixed point is based on the assumption for the stability of the spinon Fermi surface. Since it is a fixed point for critical spinon dynamics, one may investigate the stability of such a fixed point, introducing the role of critical charge fluctuations into the spin-liquid fixed point. The self-interaction constant $\lambda$ for the holon dynamics turns out to be relevant at the spin-liquid fixed point, where the upper critical dimension is $d_{c} = 7/2$. We recall that the upper critical dimension of the Yukawa coupling constant (the gauge charge) is $d_{c} = 5/2$. As a result, the self-interaction constant cannot be renormalized by their self-interactions at this fractional dimension $d_{c} = 5/2$ within the scheme of dimensional regularization. This does not mean that there do not exist renormalization effects on the self-interaction constant. Gauge-field fluctuations can lead holon quasiparticle excitations to decay into a bunch of incoherent particle-hole continuum spectra. However, we find that such effects do not occur at least in the one-loop order for holon self-energy corrections. There are screening effects for the holon self-interaction term, given by anisotropic quantum critical scaling of space and time at the U(1) spin-liquid fixed point. However, this screening is not enough to make the renormalization group flow of $\lambda$ irrelevant at least in the one-loop order, more strongly speaking, in the limit of $\varepsilon \rightarrow 0$. This run-away renormalization group flow of $\lambda$ leads us to conclude that such holon excitations remain gapped at the spin-liquid fixed point with a stable spinon Fermi surface. In other words, we fail to reach the spin-liquid to Fermi-liquid Mott critical point, given by the condensation of holons. Critical spin dynamics is given by the U(1) spin-liquid interacting fixed point as the spinon Fermi-surface problem in the absence of charge fluctuations \cite{SSL_Dimensional_Regularization_Nematic}.

In this study we focus on another possibility, giving up the stability of the spinon Fermi surface in order to describe the Mott metal-insulator transition. Since the spin-liquid to Fermi-liquid insulator-metal quantum phase transition is described by the Higgs condensation transition in the holon dynamics, it may be natural to keep the boson dynamics as a fixed-point ensemble at spin-liquid Mott quantum criticality. In other words, the relativistic holon spectrum in both $x-$ and $y-$ directions is assumed as our starting fixed point of scale invariance. Then, we find that the spinon Fermi surface cannot be stabilized at this boson fixed point, where the curvature part of the spinon spectrum becomes irrelevant and the spinon dispersion shows the one-dimensional relativistic spectrum. The spinon dynamics remains itinerant along the direction of the Fermi velocity while spinons become localized along the direction of the Fermi surface. Critical spinon dynamics at ultraviolet (UV) is given by Luttinger liquid theory \cite{Luttinger_Liquid_Review}. The effective field theory is as follows: First, critical spinons are described by the one-dimensional Dirac spectrum, coupled with U(1) gauge fluctuations. Second, critical holons are described by the two-dimensional relativistic spectrum with their self-interactions, coupled to U(1) gauge fluctuations. This critical field theory shows an emergent enhanced symmetry than that of the U(1) spin-liquid fixed point, that is, the emergent Lorentz symmetry at UV beyond the U(1) spin-liquid fixed point.

We emphasize that the rotational symmetry does not break down although the spectrum is localized along one direction. We recall that the effective field theory is represented in the double-patch construction. The double-patched effective field theory should be taken into account for all angles of the Fermi surface, where other double-patched effective field theories do not communicate with each other as discussed before, thus regarded to be independent \cite{Patch_Construction}. As a result, the rotational symmetry is preserved.

Now, it is straightforward to apply the dimensional regularization technique for the renormalization group analysis to this Lorentz-invariant critical field theory. The one-dimensional spinon Fermi surface with a flat band along the direction of the Fermi surface remains unchanged in the dimensional regularization scheme. The upper critical dimension of the Yukawa coupling constant, i.e., the gauge charge is the same as the self-interaction coupling constant, given by $d_{c} = 3$. As a result, not only the gauge coupling constant but also the self-interaction coupling constant is screened to show an interacting fixed point at IR beyond the U(1) spin-liquid fixed point discussed before. More importantly, the emergent Lorentz symmetry does not allow the appearance of the Landau damping term in the dynamics of U(1) gauge fluctuations. As a result, an interacting fixed point appears to be identified with an inverted XY universality class \cite{IXY_Review}, controlled within the dimensional regularization technique when the flavor number of holons is over its critical value. On the other hand, a fluctuation-driven first order transition \cite{Fluctuation_Driven_First_Order_I,Fluctuation_Driven_First_Order_II} results when it is below the critical number. In particular, we propose that the destabilization of a spinon Fermi surface and the emergence of one-dimensional spin dynamics near the spin-liquid Mott quantum critical point can be checked out by spin susceptibility with a $2 k_{F}$ transfer momentum, where $k_{F}$ is a Fermi momentum in the U(1) spin-liquid state: The absence of Landau damping in U(1) gauge fluctuations gives rise to a divergent behavior at zero temperature while it vanishes in the presence of a spinon Fermi surface \cite{SSL_Dimensional_Regularization_Nematic}.

Recently, emergence of localized magnetic moments from itinerant fermions has been discussed in the critical field theory of fermions and order-parameter fluctuations with their Yukawa coupling interactions \cite{Absence_Landau_Damping}. An important assumption in this renormalization group analysis is that the Landau damping term does not arise due to a certain reason, not clarified in these previous studies. The present field theoretical construction serves more transparent physical mechanism for the absence of the Landau damping term, where the emergent Lorentz symmetry plays an important role in the localization phenomenon.

\section{Effective field theory for spin-liquid Mott quantum criticality}

\subsection{U(1) slave-rotor theory for the Hubbard model}

We start from the Hubbard model as an effective Hamiltonian for $\kappa-$class organic salts \cite{Kappa_BEDT_Review}
\begin{eqnarray}
S
&=& \int_0^\beta\!d\tau \ \Bigg\{ \sum_{i} c_{i\sigma}^\dagger \big( \partial_\tau - \mu \big) c_{i\sigma} \nn &-& t\sum_{i j} \left( c_{i\sigma}^\dagger c_{j\sigma} + H.c. \right) + U \sum_{i} n_{i\uparrow} n_{i\downarrow} \Bigg\} .
\end{eqnarray}
$c_{i\sigma} = c_{i\sigma}(\tau)$ is an electron annihilation operator at site $i$ with spin $\sigma = \uparrow, \downarrow$, $n_{i\sigma} = c_{i\sigma}^\dagger c_{i\sigma}$ is electron density with spin $\sigma$, $t$ is a hopping parameter between nearest neighboring sites, $U$ is an on-site Hubbard interaction, $\mu$ is a chemical potential, and $\beta$ is an inverse temperature. The summation over the repeated spin indices is assumed in this and all expressions hereafter.

This effective Hamiltonian has an SU$_{s}$(2)$\times$SU$_{c}$(2) global symmetry at half filling, where the former and latter are involved with rotations in the spin and particle-hole spaces, respectively \cite{Hermele_SU2SR}. Here, we take into account the spin-singlet channel only, where the charge SU$_{c}$(2) symmetry is involved. Actually, interactions of both particle-hole and particle-particle channels can be incorporated to respect the SU$_{c}$(2) symmetry, realized in the SU(2) slave-rotor representation: Not only density fluctuations but also superconducting correlations are described on equal footing in the strong coupling approach \cite{KSKim_SU2SR}. In this study we focus on density fluctuations only and leave the role of the particle-particle channel in spin-liquid Mott quantum criticality as a future problem. Performing the Hubbard-Stratonovich transformation for the density-fluctuation channel, we obtain
\begin{eqnarray}
S &=& \int_0^\beta\!d\tau \ \Bigg\{ \sum_{i} c_{i\sigma}^\dagger \big( \partial_\tau - \mu + i\phi_i \big) c_{i\sigma} \nn
&-& t\sum_{ij} \left( c_{i\sigma}^\dagger c_{j\sigma} + H.c. \right) +  \frac{1}{U} \sum_i \phi_i^2 \Bigg\},
\label{Potential}
\end{eqnarray}
where $\phi_i$ is Hubbard-Stratonovich field.

A question is how to obtain a metal-insulator transition without any symmetry breaking based on this effective action. A direction would be to gap out zero sound modes without introducing local order parameters. This is completely nonperturbative. Nobody succeeded in such a nonperturbative task, starting from a Landau's Fermi-liquid state, as far as we know. An idea is to decompose an electron field as follows \cite{U1SR}: \bqa && c_{j\sigma} = e^{-i\theta_j} f_{j\sigma} . \label{U1SR} \eqa A fermion field $f_{j\sigma}$ carries only the spin quantum number $\sigma$, referred to as spinon. A boson field $\theta_{j}$ represents the conjugate variable of the density field $n_{j} = c_{j\sigma}^{\dagger} c_{j\sigma}$. Thus, their correlations reflect collective behaviors of density fluctuations. When such charged bosons are condensed, the spectrum of the $\theta_{j}$ field corresponds to that of the zero sound mode. This description is consistent with Landau's Fermi-liquid theory, where the condensation amplitude represents the quasiparticle weight. When holons become gapped, increasing the Hubbard interaction, both electron quasiparticles and zero sound modes disappear. Introducing the U(1) slave-rotor representation of Eq. (\ref{U1SR}) into Eq. (\ref{Potential}) and shifting the $\phi_j$ field as $\phi_j + \partial_\tau \theta_j$, we obtain
\begin{eqnarray}
S
&=& \int_0^\beta\!d\tau \ \Bigg\{ \sum_{i} f_{i\sigma}^\dagger \big( \partial_\tau - \mu + i\phi_i \big) f_{i\sigma} \nn
&-& t\sum_{ij} \left( f_{i\sigma}^\dagger e^{i\theta_i} e^{-i\theta_j}f_{j\sigma} + H.c. \right) \nn
&+& \frac{1}{U} \sum_i \big( \partial_\tau \theta_i + \phi_i \big)^2 \Bigg\}.
\end{eqnarray}

In order to describe the Higgs transition of the boson field, it is convenient to adopt the nonlinear $\sigma-$model field-theory approach \cite{U1SR}, replacing $e^{i \theta_{j}}$ with $b_{j}$, where the unimodular constraint of $b_{j}^{\dagger} b_{j} = 1$ should be incorporated. We write down $\partial_\tau \theta_i = -i b_i^\dagger \partial_\tau b_i$ with the following term $\int_0^\beta\! d\tau i \sum_i \lambda_i \left( b_i^\dagger b_i - 1 \right)$ to impose the rotor constraint, where $\lambda_i$ is a Lagrange multiplier field. The next step is to decompose the kinetic energy term in an appropriate way. Based on experimental results for $\kappa-$class organic salts \cite{Kappa_BEDT_Review}, we assume the presence of a spinon Fermi surface. In order to keep the existence of the spinon Fermi surface, we adopt the following ansatz for the mean-field solution \cite{SSLee_PALee_Kappa_BEDT}:
\begin{eqnarray}
&&\big< f_{i\sigma} f_{j\sigma}^\dagger \big> = -\chi^b e^{-ia_{ij}}, \ \ \
\big< b_i b_j^\dagger \big> = \chi^f e^{-ia_{ij}}
\end{eqnarray}
with $i \lambda_i = \lambda$ and $\phi_i = 0$, where the latter gives the condition of half filling. Here, we include a phase-fluctuation field $a_{ij} = a_{ij}(\tau)$ for both $\big< f_{i\sigma} f_{j\sigma}^\dagger \big>$ and $\big< b_i b_j^\dagger \big>$, which satisfies a relation $a_{ij} = -a_{ji}$.

The resulting effective theory for the insulator-metal quantum phase transition from a U(1) spin-liquid state to a Landau's Fermi-liquid phase is as follows:
\begin{eqnarray}
S
&=& \int_0^\beta\!d\tau \ \Bigg\{ \sum_{i} f_{i\sigma}^\dagger \big( \partial_\tau - \mu \big) f_{i\sigma} - t\chi^f \sum_{ij} \Big( f_{i\sigma}^\dagger e^{-ia_{ij}} f_{j\sigma} \nn &+& H.c. \Big) + \sum_{i} b_i^\dagger \left( -\frac{1}{U} \partial_\tau^2 + \lambda \right) b_i - t\chi^b \sum_{ij} \Big( b_i^\dagger e^{-ia_{ij}} b_j \nn &+& H.c. \Big) + N \left( zt\chi^f \chi^b - \lambda \right) \Bigg\} .
\end{eqnarray}
$z$ is the coordination number (e.g., $z=6$ for a triangular lattice in two dimensions), and $N$ is the total number of lattice sites. It is interesting to notice that this effective action has the following gauge symmetry:
\begin{eqnarray}
f_{i\sigma} \rightarrow e^{i\alpha_i} f_{i\sigma} , \ \ \
b_i \rightarrow e^{i\alpha_i} b_i , \ \ \ a_{ij} \rightarrow a_{ij} - \alpha_i + \alpha_j ,
\end{eqnarray}
where the phase field of the hopping parameter plays the role of the spatial component of the U(1) gauge field. Physically, such U(1) gauge fluctuations describe low lying spin-singlet fluctuations, expected to appear when excited spin states are rather ``degenerate" due to special entangled patterns of spins and such entangled dynamics gives rise to spin-singlet excitations as low-energy fluctuations instead of spin-triplet excitations. Interestingly, these spin-singlet excitations couple to density fluctuations in the way of minimal coupling, affecting critical charge dynamics seriously, compared with the case of the absence of gauge fluctuations in the mean-field level. Potential fluctuations described by $\phi_{j}$ are gapped due to the presence of a spinon Fermi surface, referred to as Debye screening. In the Coulomb gauge the temporal component decouples with the spatial part, and thus, safely ignored at low energies.

In the continuum limit we reach the following expression
\begin{eqnarray}
S
&=& \int_0^\beta\!d\tau \int\!d^2 x \ \Bigg\{ f_{\sigma}^\dagger \big( \partial_\tau - \mu - t\chi^f \vec\nabla^2 \big) f_{\sigma} \nn &+& it\chi^f \vec a \cdot \big( f_\sigma^\dagger \vec\nabla f_\sigma - \vec\nabla f_\sigma^\dagger f_\sigma \big) + t\chi^f \vec a^2 f_\sigma^\dagger f_\sigma \nn
&+& b^\dagger \left( -\frac{1}{U} \partial_\tau^2 + \lambda - t\chi^b \vec\nabla^2 \right) b + it\chi^b \vec a \cdot \big( b^\dagger \vec\nabla b - \vec\nabla b^\dagger b \big) \nn &+& t\chi^b \vec a^2 b^\dagger b + \frac{1}{4e^2} f_{\mu\nu} f_{\mu\nu} \Bigg\} + \beta N \left( zt\chi^f \chi^b - \lambda \right) . \label{EFT_Full_FS}
\end{eqnarray}
$f_\sigma = f_\sigma (\tau, \vec x)$, $b = b(\tau, \vec x)$, and $\vec a = \vec a(\tau, \vec x)$ are field variables in the continuum limit. The free part (Maxwell dynamics) of the gauge field results from the procedure of renormalization integrating over high-energy fluctuations, given by $f_{\mu\nu} = \partial_\mu a_\nu - \partial_\nu a_\mu$ ($\mu, \nu = 0, 1, 2$) with a coupling constant $e$ at a given UV scale. $\lambda$ plays the role of mass in the holon spectrum, determined self-consistently in the mean-field analysis. Neglecting U(1) gauge fluctuations in the mean-field approximation ($e \rightarrow 0$), it is straightforward to solve the resulting Gaussian-type action. One can obtain self-consistent equations for three order parameters of $\chi^{f}$, $\chi^{b}$, and $\lambda$. In $U > U_{c}$ one finds a gapped spectrum of holons given by $\lambda > 0$, where the quasiparticle weight vanishes, identified with a U(1) spin-liquid state with a spinon Fermi surface. Decreasing the Hubbard interaction, $\lambda$ becomes also reduced to touch zero at $U = U_{c}$, identified with a Mott critical point in the mean-field analysis. Further reduction of $U$ does not change $\lambda = 0$ in $U < U_{c}$. But, the holon condensation should occur in order to satisfy the rotor constraint of $\langle b^{\dagger}(\tau, \vec x) b(\tau, \vec x) \rangle = 1$, giving rise to finite quasiparticle weight in the electron spectrum and recovering a Landau's Fermi-liquid state. In this study we discuss how this mean-field structure is modified beyond the mean-field approximation, introducing the role of U(1) gauge fluctuations and $\lambda-$field fluctuations in the spin-liquid Mott transition.

\subsection{Effective field theory for spin-liquid Mott quantum criticality}

\begin{figure}[t]
\includegraphics[width=9cm]{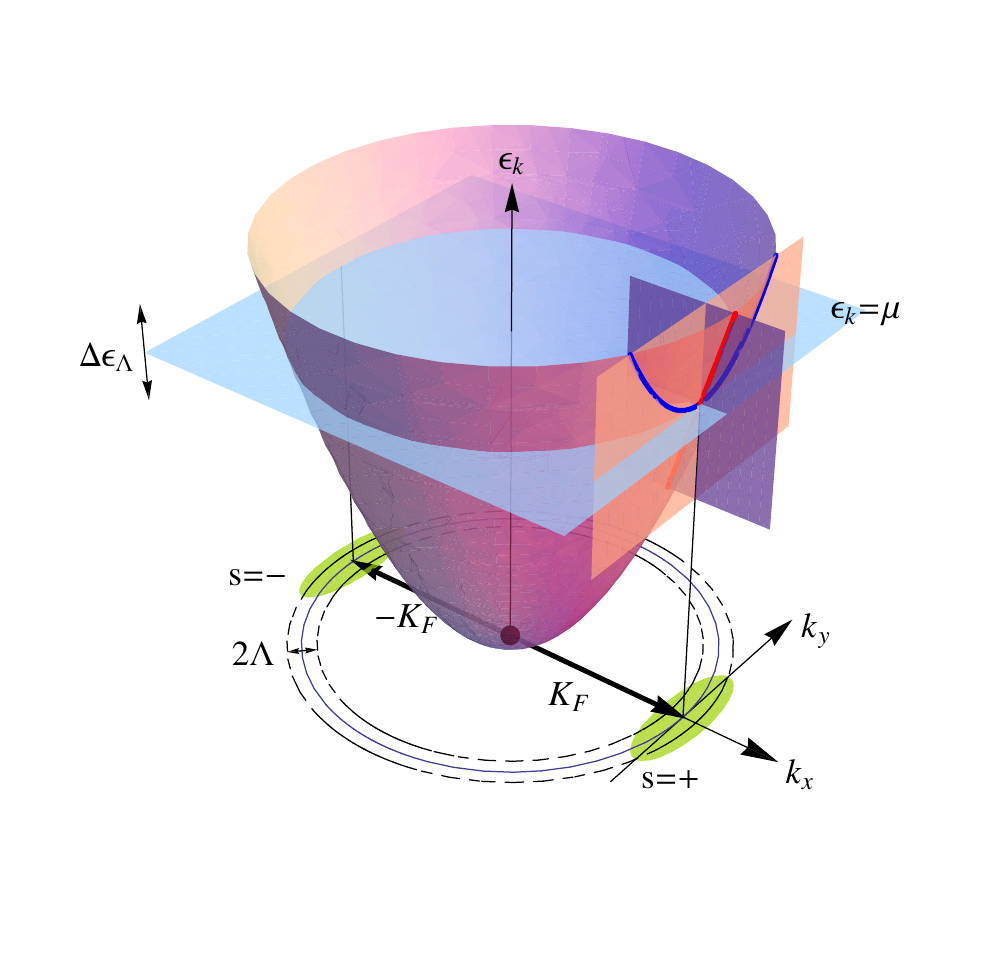}
\caption{Patch construction for a two-dimensional Fermi-surface problem. The paraboloid represents spinon dispersion, the red and the blue lines are the dispersion along the $k_x$ and $k_y$, respectively. In the minimal model, only two patches on the opposite side of the Fermi surface (the shaded regions in $k_x k_y$-plane) are considered.
}
\label{fig:patch}
\end{figure}

\begin{figure}[t]
\includegraphics[width=5cm]{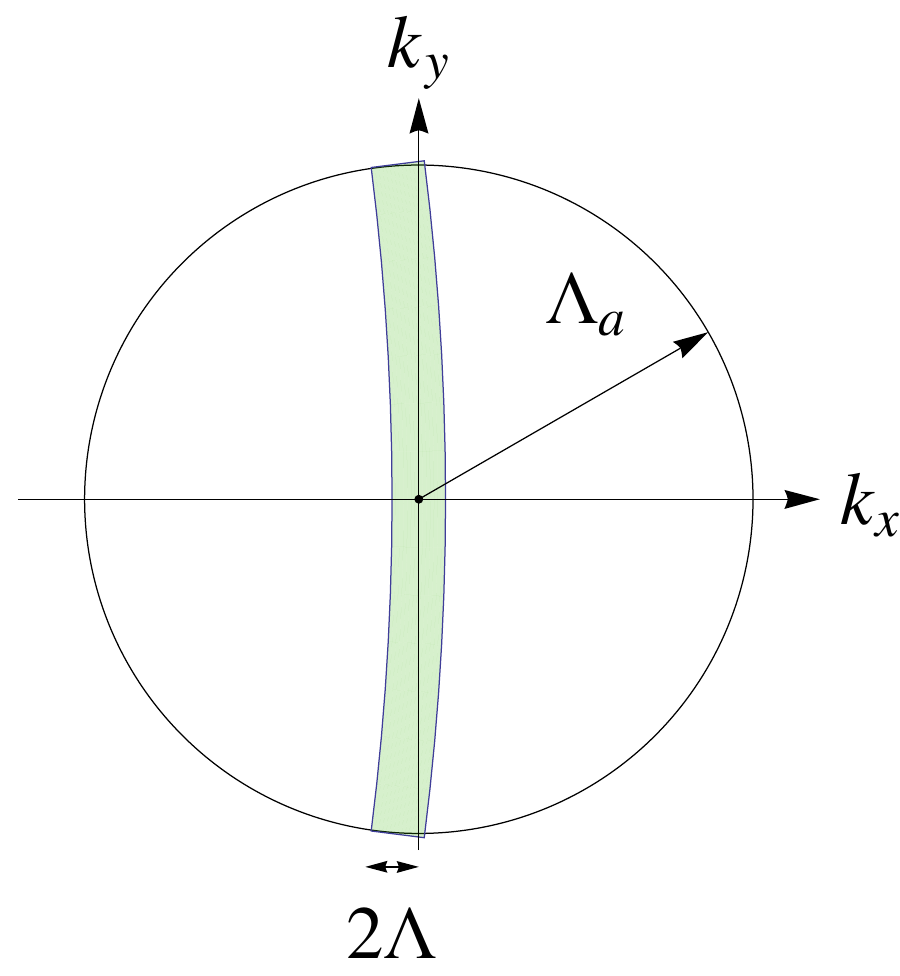}
\caption{Cutoffs in a two-dimensional Fermi-surface problem. $\Lambda$ and $\Lambda_a$ are cutoffs of spinon and gauge field, respectively.}
\label{fig:int_range}
\end{figure}

We construct an effective field theory in the double-patch description, regarded to be a minimal model for spin-liquid Mott quantum criticality and justified in the IR limit since communications between different patches are renormalization-group irrelevant \cite{SSL_Breakdown_N_Before,Patch_Construction}. When linearizing the spinon dispersion perpendicular to the Fermi surface, we have
\begin{eqnarray}
-i\omega - \mu + t\chi^f \vec k^2
\rightarrow -i\omega + v_F k_x + t\chi^f k_y^2 ,
\end{eqnarray}
where $v_F = 2t\chi^f K_F$ is a Fermi velocity, $K_F$ is a Fermi wave vector, and the chemical potential $\mu$ is tuned to give half filling. Here, the momentum is redefined from the Fermi surface. Then, Eq. (\ref{EFT_Full_FS}) can be written as follows in the double-patch construction
\begin{eqnarray}
&&S = S_{f} + S_{b} + S_{a} + S_{fa} + S_{ba}, \nn
&&S_{f} = \int_k \ f_{\sigma s}^\dagger (k) \big( ik_0 + sv_F k_x + t\chi^f k_y^2 \big) f_{\sigma s} (k), \nn
&&S_{b} = \int_k \ b^\dagger(k) \left( \frac{1}{U} k_0^2 + t\chi^b \vec k^2 + \lambda \right) b(k), \nn
&&S_{a} = \frac{1}{2} \int_q \ a(-q) \big( q_0^2 + \vec q^2 \big) a(q), \nn
&&S_{fa} = -v_F e \int_{k, q} \ s a(q) f_{\sigma s}^\dagger (k+q) f_{\sigma s}(k) \nn
&&+ t\chi^f e^2 \int_{k,p,q} \ \vec a(-p+q) \cdot \vec a(p) f^\dagger_{\sigma s}(k+q) f_{\sigma s}(k), \nn
&&S_{ba} = -2t\chi^b e \int_{k,q} \vec k \cdot \vec a(q) b^\dagger(k+q) b(k) \nn
&&+ t\chi^b e^2 \int_{k,p,q} \vec a(-p+q) \cdot \vec a(p) b^\dagger (k+q) b(k).
\end{eqnarray}
See Fig. \ref{fig:patch}. We rescaled the gauge field as $\vec a \rightarrow e \vec a$ and abbreviated the integral as $\int_k = \int d^3k/(2\pi)^3$. $s=\pm$ is a patch index. We note that the integration region in the gauge-spinon field coupling, $S_{fa}$, is given by a narrow strip as shown in Fig. \ref{fig:int_range}.

Following D. Dalidovich and S.-S. Lee \cite{SSL_Dimensional_Regularization_Nematic}, we introduce a Dirac spinor for dimensional regularization, combining the two spinon fields of opposite patches as
\begin{eqnarray}
\psi_\sigma(k) = \begin{pmatrix} f_{\sigma +}(k) \\ f^\dagger_{\sigma -}(-k) \end{pmatrix} .
\end{eqnarray}
Then, the spinon part of the above effective action can be written in the form of the $(1+1)$-dimensional Dirac theory
\begin{eqnarray}
&&S_{f} = \int_k \bar \psi_\sigma (k) \big( i\gamma_0 k_0 + i\gamma_1 \delta_k \big) \psi_\sigma(k) , \nn
&&S_{fa} = iv_Fe \int_{k,q} a(q) \bar\psi_\sigma(k+q) \gamma_5 \gamma_1 \psi_\sigma(k),
\end{eqnarray}
where $\gamma_0 = \sigma_y$, $\gamma_1 = \sigma_x$ , $\gamma_5 = i\gamma_0 \gamma_1 = \sigma_z$, $\delta_k = v_F k_x + t\chi^f k_y^2$ is the dispersion of spinons near the Fermi surface, and $\bar\psi = \psi^\dagger \gamma_0$. Although the second term in $S_{fa}$ is not shown explicitly here, its role is incorporated to preserve the U(1) gauge symmetry in the renormalization group analysis.

Now, it is straightforward to construct the setup for dimensional regularization. Extending the co-dimension of the spinon Fermi surface, we obtain
\begin{eqnarray}
&&S = S_f + S_b + S_a + S_{fa} + S_{ba}, \nn
&&S_f = \int_k \bar\psi_\sigma (k) \big( i{\bm \Gamma} \cdot {\bm K} + i\gamma_{d-1} \delta_k \big) \psi_\sigma(k), \nn
&&S_b = \int_k b_a^\dagger(k) \left( \frac{1}{U} {\bm K}^2 + t\chi^b \vec k^2 \right) b_a(k), \nn
&&+ \frac{\lambda}{4N} \int_{k,p,q} b_a^\dagger(k+q) b_a(k) b_b^\dagger(p-q) b_b(p) \nn
&&S_a = \frac{1}{2} \int_q a(-q) \big( {\bm Q}^2 + \vec q^2 \big) a(q), \nn
&&S_{fa} = \frac{iv_F e}{\sqrt{N}} \int_{k,q} a(q) \bar\psi_\sigma(k+q) \gamma_5 \gamma_{d-1} \psi_\sigma(k), \nn
&&S_{ba} = -\frac{2t\chi^be}{\sqrt{N}} \int_{k,q} \vec k \cdot \vec a(q) b_a^\dagger(k+q) b_a(k) \nn
&&+ \frac{t\chi^b e^2}{N} \int_{k,p,q} \vec a(-p+q) \cdot \vec a(p) b_a^\dagger(k+q) b_a(k).
\label{eq:DimRegAction2}
\end{eqnarray}
Here, $\int_k$ is now a ($d+1$)-dimensional integral $\int_k = \int d^{d+1}k / (2\pi)^{d+1}$. The bold-faced vector ${\bm K}$ is a ($d-1$)-dimensional vector ${\bm K} = (k_0, k_1, \cdots, k_{d-2})$, and the arrowed vector $\vec k$ is a $2$-dimensional vector $\vec k = (k_{d-1}, k_d)$. $\delta_k = v_F k_{d-1} + t\chi^f k_d^2$ is an equipotential surface near the Fermi surface. The unimodular constraint of $b^\dagger b = 1$ is softened by the $b^4$-interaction term with $\lambda$. In particular, we point out that the flavor number of holons shown in the subscript of $b_{a}(k)$ is generalized from $a = 1$ to $a = 1, 2, \dots, N$. It is natural to consider that the number of holon flavors control the strength of quantum fluctuations, for example, screening of effective interactions.

\section{Renormalization group analysis I: A spin-liquid fixed point with a stable spinon Fermi surface}
\label{Sec:RG_I}

In this section we perform the renormalization group analysis, based on the stability of the spinon Fermi surface. It is convenient to simplify the spinon sector, rescaling both fields and coupling constants:
\begin{eqnarray}
&&{\bm K} \rightarrow v_F {\bm K}, \nn
&&\psi_\sigma \rightarrow \psi_\sigma / v_F^{d\over2}, \ \ \
b \rightarrow b/ (v_F^{d-1} t\chi^b)^{1\over2}, \nn
&&a \rightarrow a/v_F^{d-1\over2}, \nn
&&e \rightarrow e/v_F^{d-1\over2}, \ \ \
\lambda \rightarrow \lambda (t\chi^b)^2/v_F^{d-1}.
\end{eqnarray}
Now, the effective action can be written as
\begin{eqnarray}
&&S = S_f + S_b + S_a + S_{fa} + S_{ba}, \nn
&&S_f = \int_k \bar\psi_\sigma (k) \big( i{\bm \Gamma} \cdot {\bm K} + i\gamma_{d-1} \delta_k \big) \psi_\sigma(k), \nn
&&S_b = \int_k b_a^\dagger(k) \left( \zeta_b^2 {\bm K}^2 + \vec k^2 \right) b_a(k) \nn
&&+ \frac{\lambda}{4N} \int_{k,p,q} b_a^\dagger(k+q)b_a(k) b_b^\dagger(p-q) b_b(p), \nn
&&S_a = \frac{1}{2} \int_q a(-q) \left( \zeta_a^2 {\bm Q}^2 + \vec q^2 \right) a(q), \nn
&&S_{fa} = \frac{ie}{\sqrt{N}} \int_{k,q} a(q) \bar\psi_\sigma(k+q) \gamma_5 \gamma_{d-1} \psi_\sigma(k), \nn
&&S_{ba} = -\frac{2e}{\sqrt{N}} \int_{k,q} \vec k \cdot \vec a(q) b_a^\dagger(k+q) b_a(k)\nn
&&+ \frac{e^2}{N} \int_{k,p,q} \vec a(-p+q) \cdot \vec a(p) b_a^\dagger(k+q) b_a(k),
\label{eq:DimRegAction}
\end{eqnarray}
where $\delta_k = k_{d-1} + \kappa k_d^2$, $\kappa = t\chi^f/v_F$, $\zeta_b^2 = v_F^2/(Ut\chi^f)$, and $\zeta_a^2 = v_F^2$. We choose the unit such that $\kappa =1$.

\subsection{Scaling analysis: Multiple interaction-energy scales}

As discussed in the introduction, this Fermi-surface fixed point turns out to be too stable to allow a metal-insulator transition within the perturbative renormalization group analysis. We consider the scaling transformation that preserves the spinon dispersion:
\begin{eqnarray}
&&{\bm K} = \frac{{\bm K}'}{s}, \ \ \
k_{d-1} = \frac{k_{d-1}'}{s}, \ \ \
k_d = \frac{k_d'}{\sqrt{s}} .
\end{eqnarray}
Here, $s$ is now a scaling factor, not the path index as in previous section. In order to make the spinon sector invariant under this scaling transformation, we introduce \bqa && \psi_\sigma(k) = s^{\Delta_\psi} \psi'_\sigma(k') \eqa into $S_{f}$ in Eq. (\ref{eq:DimRegAction}), and obtain $\Delta_\psi = \frac{d}{2} + \frac{3}{4}$. For the free part of the gauge field, only the $q_d^2$-term is marginal and others are irrelevant when the gauge field scales
as
\begin{eqnarray}
a(q) = s^{\Delta_a} a'(q'), \ \ \
\Delta_a = \frac{d}{2} + \frac{3}{4} ,
\end{eqnarray}
the same as the scaling dimension $\Delta_{\psi}$ of the spinon field. Then, the gauge-coupling $e$ scales as
\begin{eqnarray}
e = s^{\Delta_e} e', \ \ \
\Delta_e = \frac{d}{2} - \frac{5}{4} ,
\end{eqnarray}
read from the spinon-gauge field coupling. The gauge charge $e$ is relevant in $d<{5 \over 2}$, irrelevant in $d>{5\over2}$, and marginal at $d={5\over2}$. We concentrate on the dimension $d={5\over2} - \epsilon$ with small $\epsilon>0$ for the controllable renormalization group analysis in the $\epsilon-$expansion.

The scale transformation of the boson field $b$ is not independent, since the coupling constant $e$ of the boson-gauge field vertex should be the same as that of the spinon-gauge field vertex, resulting from the gauge symmetry. Therefore, we have the scaling transformation
\begin{eqnarray}
b_a(k) = s^{\Delta_b} b'_a(k'), \ \ \
\Delta_b = \frac{d}{2} + \frac{5}{4}.
\end{eqnarray}
As a result, both ${\bm K}^2$-term and $k_{d-1}^2$-term in the free-part of the boson field are marginal, while the $k_d^2$-term is relevant, regardless of the dimension. This $b$-field scaling leads the self-interaction $\lambda$ to scale as
\begin{eqnarray}
\lambda = s^{\Delta_\lambda} \lambda', \ \ \
\Delta_\lambda = d-\frac{7}{2},
\end{eqnarray}
which is relevant in $d={5\over2}-\epsilon$.

The above scaling analysis gives us the renormalized effective field theory
\begin{eqnarray}
&&S = S_f + S_b + S_a + S_{fa} + S_{ba}, \nn
&&S_f = \int_k \bar\psi_\sigma (k) \big( i{\bm \Gamma} \cdot {\bm K} + i\gamma_{d-1} \delta_k \big) \psi_\sigma(k), \nn
&&S_b = \int_k b_a^\dagger(k) \left( \zeta_b^2 {\bm K}^2 + \vec k^2 \right) b_a(k) \nn
&&+ \frac{\lambda\mu^{1+\epsilon}}{4N} \int_{k,p,q} b_a^\dagger(k+q) b_a(k) b_b^\dagger(p-q) b_b(p), \nn
&&S_a = \frac{1}{2} \int_q q_d^2 a(-q) a(q), \nn
&&S_{fa} = \frac{ie\mu^{\epsilon\over2}}{\sqrt{N}} \int_{k,q} a(q) \bar\psi_\sigma(k+q) \gamma_5 \gamma_{d-1} \psi_\sigma(k), \nn
&&S_{ba} = - \frac{2e\mu^{\epsilon\over2}}{\sqrt{N}} \int_{k,q} \vec k \cdot \vec a(q) b_a^\dagger(k+q) b_a(k)\nn
&&+ \frac{e^2 \mu^{\epsilon}}{N} \int_{k,p,q} \vec a(-p+q) \cdot \vec a(p) b_a^\dagger(k+q) b_a(k) , \label{Renormalized_EFT}
\end{eqnarray}
where we introduced the parameter $\mu$ identified with a mass scale (not the chemical potential as in the previous section), and dropped the irrelevant terms in $S_a$.

\subsection{Renormalization group analysis}

For the renormalization group analysis, we rewrite the effective bare action in terms of bare field variables and coupling parameters as the renormalized effective action and counter terms in terms of renormalized field variables and interaction parameters:
\bqa && S_{B} = S + S_{CT} , \eqa
where the bare action is given by
\bqa
&& S_B
= \int_{k_B} \bar\psi_{B\sigma} (k_B) \big( i{\bm \Gamma} \cdot {\bm K}_B + i\gamma_{d-1} \delta_{k_B} \big) \psi_{B\sigma}(k_B)
\nn
&& + \int_{k_B} b_{Ba}^\dagger(k_B) \left( \zeta_{bB}^2 {\bm K}_B^2 + \vec k_B^2 \right) b_{Ba}(k_B) \nn &&
+ \frac{\lambda_B}{4N} \int_{k_B,p_B,q_B} b_{Ba}^\dagger(k_B+q_B) b_{Ba}(k_B) \nn && b_{Bb}^\dagger(p_B-q_B) b_{Bb}(p_B) \nn
&& + \frac{1}{2} \int_{q_B} q_{Bd}^2 a_B(-q_B) a(q_B) \nn
&& + \frac{ie_B}{\sqrt{N}} \int_{k_B,q_B} a_B(q_B) \bar\psi_{B\sigma}(k_B+q_B) \gamma_5 \gamma_{d-1} \psi_{B\sigma}(k_B) \nn
&&-\frac{2e_B}{\sqrt{N}} \int_{k_B,q_B} \vec k_B \cdot \vec a_B(q_B) b_{Ba}^\dagger(k_B+q_B) b_{Ba}(k_B) \nn &&
+ \frac{e_B^2}{N} \int_{k_B,p_B,q_B} \vec a_B(-p_B+q_B) \cdot \vec a_B(p_B) \nn
&& b_{Ba}^\dagger(k_B+q_B) b_{Ba}(k_B)
\eqa
and the counter terms are described by
\bqa
&& S_{CT} = \int_k \bar\psi_\sigma (k) \big( A_{\psi 1} i{\bm \Gamma} \cdot {\bm K} + A_{\psi 2} i\gamma_{d-1} \delta_k \big) \psi_\sigma(k)
\nn && + \int_k b_a^\dagger(k) \left( A_{b1} \zeta_b^2 {\bm K}^2 + A_{b2} \vec k^2 \right) b_a(k)
\nn && + A_\lambda \frac{\lambda\mu^{1+\epsilon}}{4N} \int_{k,p,q} b_a^\dagger(k+q) b_a(k) b_b^\dagger(p-q) b_b(p)
\nn && + A_{a2} \frac{1}{2} \int_q q_d^2 a(-q) a(q)
\nn && + A_{\psi a} \frac{ie\mu^{\epsilon\over2}}{\sqrt{N}} \int_{k,q} a(q) \bar\psi_\sigma(k+q) \gamma_5 \gamma_{d-1} \psi_\sigma(k)
\nn && -A_{ba1} \frac{2e\mu^{\epsilon\over2}}{\sqrt{N}} \int_{k,q} \vec k \cdot \vec a(q) b_a^\dagger(k+q) b_a(k)
\nn && + A_{ba2} \frac{e^2 \mu^{\epsilon}}{N} \int_{k,p,q} \!\!\!\! \vec a(-p+q) \cdot \vec a(p) b_a^\dagger(k+q) b_a(k).
\eqa
The renormalized effective action is given by Eq. (\ref{Renormalized_EFT}). We note that the term with $\zeta_b$ is allowed to flow.

The relation between bare and renormalized quantities are
\begin{eqnarray}
&&{\bm K} = \frac{Z_{\psi 2}}{Z_{\psi 1}} {\bm K}_B , \ \ \
\vec k = \vec k_B, \nn
&&\psi_{\sigma}(k) = Z_\psi^{-{1\over2}} \psi_{B\sigma}(k_B), \ \ \
Z_\psi = Z_{\psi 2} \left( \frac{Z_{\psi 2}}{Z_{\psi 1}} \right)^{d-1}, \nn
&&b_a(k) = Z_{b}^{-{1\over2}} b_{Ba}(k_B), \ \ \
Z_{b} = Z_{b2} \left( \frac{Z_{\psi 2}}{Z_{\psi 1}} \right)^{d-1}, \nn
&&a(q) = Z_{a}^{-{1\over2}} a_B(q_B), \ \ \
Z_{a} = Z_{a2} \left( \frac{Z_{\psi 2}}{Z_{\psi 1}} \right)^{d-1}, \nn
&&e_B = e\mu^{\epsilon\over2} Z_{a2}^{-{1\over2}} \left( \frac{Z_{\psi 2}}{Z_{\psi 1}} \right)^{d-1\over2}, \nn
&&\lambda_B = \lambda \mu^{1+\epsilon} Z_\lambda Z_{b2}^{-2} \left( \frac{Z_{\psi 2}}{Z_{\psi 1}} \right)^{d-1}, \nn
&&\zeta_{bB}^2 = \zeta_b^2 \frac{Z_{b1}}{Z_{b2}} \left( \frac{Z_{\psi 2}}{Z_{\psi 1}} \right)^{2} ,
\end{eqnarray}
where $Z_{i} = 1+ A_{i}$ are renormalization constants with $i=\psi 1, \psi 2, b1, b2, a2, \psi e, ba1, ba2, \lambda$. Here, we used the Ward identity,
\begin{eqnarray}
Z_{\psi 2} = Z_{\psi e}, \ \ \
Z_{b2} = Z_{ba1} = Z_{ba2}.
\end{eqnarray}

It is straightforward to perform the renormalization group analysis, as shown below. An essential point beyond the previous study is that there exist multiple interaction-energy scales, whose upper critical dimensions differ from each other. The upper critical dimension of the gauge-field coupling is $d_{c}^{g} = 5/2$ while that of the self-interaction parameter of the Higgs field is $d_{c}^{\lambda} = 7/2$. As a result, the self-interaction parameter is relevant in $d={5\over2}-\epsilon$, discussed before. It is almost obvious to expect the screening effect from holon fluctuations. However, it turns out that the screening effect cannot occur. The $1/\epsilon$ pole or logarithmic divergence from quantum corrections does not appear in such a fractional dimension for the renormalization of the self-interaction parameter, just originating from the property of the Gamma function. As a result, we obtain $Z_\lambda = 1$. Of course, this does not mean that there do not exist renormalization effects on the self-interaction constant. Gauge-field fluctuations can lead holon quasiparticle excitations to decay into a bunch of incoherent particle-hole continuum spectra, described by $Z_{b2}$. However, we find that such effects do not occur at least in the one-loop order for holon self-energy corrections. There are screening effects for the holon self-interaction term, given by anisotropic quantum critical scaling of space and time at the U(1) spin-liquid fixed point. However, this screening is not enough to make the renormalization group flow of $\lambda$ irrelevant at least in the one-loop order, more strongly speaking, in the limit of $\varepsilon \rightarrow 0$. Although we believe that this nonrenormalization of the self-interaction parameter is an artifact of the dimensional regularization, we do not exclude the possibility that it can be fundamental, guaranteeing the stability of the spinon Fermi surface. Since the self-interaction parameter flows to infinity in this ansatz, holons remain gapped due to such strong correlations. In other words, we fail to reach the spin-liquid to Fermi-liquid Mott critical point, given by the condensation of holons. Critical spin dynamics is given by the U(1) spin-liquid interacting fixed point just as the spinon Fermi-surface problem in the absence of charge fluctuations \cite{SSL_Dimensional_Regularization_Nematic}.

\begin{figure}[t]
\includegraphics[width=8cm]{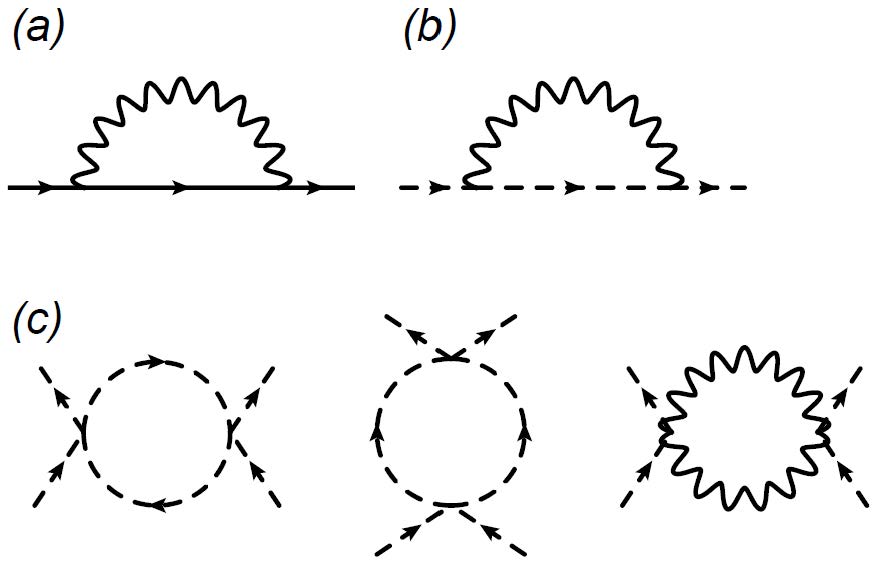}
\caption{Self-energy corrections of spinons (a) and holons (b) and $\lambda$-vertex corrections (c) in the one-loop order. The thick line represents the spinon propagator, and the dashed line describe the holon Green's function. The wavy line gives the gauge-field propagator.}
\label{fig:self_energy_RG1}
\end{figure}

Renormalization group equations for both coupling constants of $e$ and $\lambda$ result from $\mu de_B / d\mu = 0$ and $\mu d\lambda_B / d\mu = 0$, given by
\begin{eqnarray}
\beta_e
&\equiv& \mu\frac{de^2}{d\mu}
= e^2 \bigg[ - \epsilon + \frac{\mu}{Z_{a2}} \frac{d Z_{a2}}{d\mu} \nn
&&\hspace{50pt} - (d-1) \frac{\mu}{Z_{\psi 2} / Z_{\psi 1}} \frac{d Z_{\psi 2}/Z_{\psi 1}}{d\mu} \bigg], \nn
\beta_\lambda
&\equiv& \mu\frac{d\lambda}{d\mu}
= \lambda \bigg[ -(1+\epsilon) - \frac{\mu}{Z_\lambda} \frac{dZ_\lambda}{d\mu} + 2 \frac{\mu}{Z_{b2}} \frac{dZ_{b2}}{d\mu} \nn
&&\hspace{50pt} - (d-1) \frac{\mu}{Z_{\psi 2} / Z_{\psi 1}} \frac{d Z_{\psi 2}/Z_{\psi 1}}{d\mu} \bigg].
\end{eqnarray}
The renormalization constants of $Z_{\psi 1}$, $Z_{\psi 2}$, $Z_{b2}$, and $Z_{a2}$ are obtained from the self-energy corrections of spinons [Fig. \ref{fig:self_energy_RG1} (a)], holons, and gauge fields [Fig. \ref{fig:self_energy_RG1} (b)], respectively. For the evaluation of these self-energies, see Appendix \ref{App:polarization1}. Recall $Z_\lambda = 1$, given by vertex corrections [Fig. \ref{fig:self_energy_RG1} (c)]. As shown in the Appendix \ref{App:polarization1}, the $1/\epsilon$-divergence turns out to be absent in both holon and gauge-field self-energies. For the spinon self-energy, only the $Z_{\psi 2}$ constant has the $1/\epsilon$-divergence. As a result, the $\beta-$function for the gauge coupling $e$ is given by
\begin{eqnarray}
\beta_e
&=& e^2\left( - \epsilon + \frac{3-2\epsilon}{3} u_1 e^{4\over3} \right),
\end{eqnarray}
where $u_1\approx 0.0625$.
Therefore, we have an unstable fixed point at $e=0$ and a stable fixed point at $e=e_*$, given by
\begin{eqnarray}
e_*^{4\over3} = \frac{3\epsilon}{u_1 (3-2\epsilon)}.
\end{eqnarray}
At this stable fixed point, the $\beta-$function for the self-interaction constant $\lambda$ becomes
\begin{eqnarray}
\beta_\lambda^* &=& -(1+\epsilon) \lambda + \frac{3-2\epsilon}{3} u_1 e_*^{4\over3} \lambda = -\lambda.
\end{eqnarray}
This shows that the self-interaction constant of holons is not screened at all up to the one-loop level, irrespective to the dimension.

Let us summarize the renormalization group analysis for the metal-insulator transition from a Landau's Fermi-liquid state to a U(1) spin-liquid phase with a spinon Fermi surface, assuming the stability of the spinon Fermi surface. We found that the U(1) spin-liquid fixed point with a spinon Fermi surface is too stable to allow critical charge fluctuations. As a result, we cannot reach the Mott quantum critical point, where the nature of critical spinon dynamics remains essentially the same as that of the U(1) spin-liquid state with a finite fixed-point gauge coupling constant.

\section{Renormalization group analysis II: A spin-liquid Mott quantum critical point}
\label{Sec:RG_II}

\subsection{Scaling analysis: Emergent Luttinger-liquid dynamics of spinons at UV}

In this section we perform the renormalization group analysis, based on the stability of the holon dynamics. The relativistic spectrum of holons makes the curvature effect of the spinon Fermi surface become irrelevant, giving rise to localization along the direction of the Fermi surface. As a result, the spinon dynamics is described by the Luttinger-liquid spectrum at UV.

We consider the scaling transformation that preserves the holon dispersion:
\begin{eqnarray}
&&{\bm K} = \frac{{\bm K}'}{s}, \ \ \
\vec k = \frac{\vec k'}{s} . \label{Lorentz_Scaling}
\end{eqnarray}
Taking into account the scaling transformation for the holon field as
\bqa && b_a(k) = s^{\Delta_b} b_a'(k') , \eqa
we obtain $\Delta_b = \frac{d+3}{2}$.
Accordingly, the scaling transformation of the self-interaction parameter is given by
\begin{eqnarray}
\lambda = s^{\Delta_\lambda}\lambda', \ \ \
\Delta_\lambda = d-3.
\end{eqnarray}
The gauge field follows essentially the same scaling relation as $b$-field,
\begin{eqnarray}
a(q) = s^{\Delta_a} a'(q'), \ \ \
\Delta_a = \frac{d+3}{2}.
\end{eqnarray}
Then, the coupling constant $e$ scales as
\begin{eqnarray}
e = s^{\Delta_e} e', \ \ \
\Delta_e = \frac{d-3}{2}.
\end{eqnarray}
Both the self-interaction and gauge-interaction parameters are relevant in $d<3$, irrelevant in $d>3$, and marginal in $d=3$, identified with the upper critical dimension in this ansatz. We concentrate on the dimension $d=3-\epsilon$.

The scaling transformation of the spinon field from the spinon-gauge field coupling is given by
\begin{eqnarray}
\psi_\sigma(k) = s^{\Delta_\psi} \psi_\sigma'(k'), \ \ \
\Delta_\psi = \frac{d}{2} +1.
\end{eqnarray}
The scaling transformation of Eq. (\ref{Lorentz_Scaling}) leads the $k_d^2$-term in the free part of the spinon dynamics irrelevant. The dynamics of spinons becomes localized along the direction of the Fermi surface. Therefore, we start with the following effective action:
\begin{eqnarray}
&&S = S_f + S_b + S_a + S_{fa} + S_{ba}, \nn
&&S_f
= \int_{k} \bar\psi_\sigma(k) \big( i {\bm \Gamma} \cdot {\bm K} + iv_F\gamma_{d-1} k_{d-1} \big) \psi_\sigma(k), \nn
&&S_b
= \int_k b_a^\dagger(k) \left( \frac{1}{U} {\bm K}^2 + t\chi^b \vec k^2 \right) b_a(k) \nn && + \frac{\lambda\mu^\epsilon}{4N} \int_{k,p,q} b_a^\dagger(k+q)b_a(k) b_b^\dagger(p-q) b_b(p), \nn
&&S_a
= \frac{1}{2} \int_q a(-q) \big( {\bm Q}^2 + \vec q^2 \big) a(q), \nn
&&S_{fa}
= \frac{iv_Fe\mu^{\epsilon\over 2}}{\sqrt{N}} \int_{k,q} a(q) \bar\psi_\sigma(k+q) \gamma_5 \gamma_{d-1} \psi_\sigma(k), \nn
&&S_{ba}
= - \frac{2 t\chi^b e\mu^{\epsilon\over2}}{\sqrt{N}} \int_{k,q} \vec k \cdot \vec a(q) b_a^\dagger(k+q) b_a(k) \nn &&
+ \frac{t\chi^b e^2 \mu^\epsilon}{N} \int_{k,p,q} \!\!\!\vec a(-p+q) \cdot \vec a(p) b_a^\dagger(k+q) b_a(k).
\end{eqnarray}
Here, we introduced the mass parameter $\mu$. Rescaling momenta, fields, and couplings as
\bqa && {\bm K} \rightarrow \sqrt{t\chi^b U} {\bm K} , \nn && b\rightarrow b/\left[ (t\chi^bU)^{d-1\over2} t\chi^b \right]^{1\over2}, ~~~ \psi \rightarrow \psi /\left[ (t\chi^bU)^{d-1\over2} v_F \right]^{1\over2}, \nn && a \rightarrow a/ (t\chi^bU)^{d-1\over4}, \nn && e\rightarrow e/(t\chi^bU)^{d-1\over4}, ~~~ \lambda \rightarrow (t\chi^bU)^{d-1\over2} (t\chi^b)^4 \lambda ,
\eqa
we reach the following expression as our starting point:
\begin{eqnarray}
&&S = S_f + S_b + S_a + S_{fa} + S_{ba}, \nn
&&S_f
= \int_{k} \bar\psi_\sigma(k) \big( i \zeta_\psi {\bm \Gamma} \cdot {\bm K} + i\gamma_{d-1} k_{d-1} \big) \psi_\sigma(k), \nn
&&S_b
= \int_k b_a^\dagger(k) \big( {\bm K}^2 + \vec k^2 \big) b_a(k) \nn && + \frac{\lambda\mu^\epsilon}{4N} \int_{k,p,q} b_a^\dagger(k+q) b_a(k) b_b^\dagger(p-q) b_b(p), \nn
&&S_a
= \frac{1}{2} \int_q a(-q) \big( \zeta_a^2 {\bm Q}^2 + \vec q^2 \big) a(q), \nn
&&S_{fa}
= \frac{ie\mu^{\epsilon\over 2}}{\sqrt{N}} \int_{k,q} a(q) \bar\psi_\sigma(k+q) \gamma_5 \gamma_{d-1} \psi_\sigma(k), \nn
&&S_{ba}
= - \frac{2 e\mu^{\epsilon\over2}}{\sqrt{N}} \int_{k,q} \vec k \cdot \vec a(q) b_a^\dagger(k+q) b_a(k) \nn
&& + \frac{e^2 \mu^\epsilon}{N} \int_{k,p,q} \vec a(-p+q) \cdot \vec a(p) b_a^\dagger(k+q) b_a(k).
\end{eqnarray}

We would like to emphasize that the spinon Fermi surface may not be stable during the Higgs transition. Indeed, as long as the Lorentz-invariant holon spectrum or the relativistic spectrum of the zero sound mode is preserved across the Landau Fermi-liquid to U(1) spin-liquid transition, we confirmed that the spinon Fermi surface cannot be stabilized. The spinon dynamics is given by QED$_{2}$ (quantum electrodynamics in $(1+1)-$dimensions) and the holon dynamics is described by Abelian Higgs model in $(2+1)-$dimensions. As a result, the effective field theory shows an enhanced emergent symmetry at UV near the spin-liquid Mott quantum critical point.

\subsection{Renormalization group analysis}

\subsubsection{Renormalized effective action and counter terms}

For the renormalization group analysis, we rewrite the effective bare action in terms of bare field variables and coupling parameters as the renormalized effective action and counter terms in terms of renormalized field variables and interaction parameters. Recall $S_{B} = S + S_{CT}$, where the bare action is given by
\begin{eqnarray}
&& S_B = \int_{k_B} \bar\psi_{B\sigma}(k_B) \big( i \zeta_{\psi B} {\bm \Gamma} \cdot {\bm K}_B + i\gamma_{d-1} k_{d-1 B} \big) \psi_{B\sigma}(k_B) \nn
&&+ \int_{k_B} b_{Ba}^\dagger(k_B) \big( {\bm K}_B^2 + \vec k_B^2 \big) b_{Ba}(k_B) \nn && + \frac{\lambda_B}{4N} \int_{k_B,p_B,q_B} \hspace{-10pt} b_{Ba}^\dagger(k_B+q_B) b_{Ba}(k_B) b_{Bb}^\dagger(p_B-q_B) b_{Bb}(p_B) \nn && + \frac{1}{2} \int_{q_B} a_B(-q_B) \big( \zeta_{aB}^2 {\bm Q}_B^2 + \vec q_B^2 \big) a_B(q_B), \nn
&&+ \frac{ie_B}{\sqrt{N}} \int_{k_B,q_B} a_B(q_B) \bar\psi_{B \sigma}(k_B+q_B) \gamma_5 \gamma_{d-1} \psi_{B\sigma}(k_B) \nn
&&- \frac{2e_B}{\sqrt{N}} \int_{k_B,q_B} \vec k_B \cdot \vec a_B(q_B) b_{Ba}^\dagger(k_B+q_B) b_{Ba}(k_B) \nn &&
+ \frac{e_B^2}{N} \int_{k_B,p_B,q_B} \vec a_B(-p_B+q_B) \cdot \vec a_{B}(p_B) \nn&& b_{Ba}^\dagger(k_B+q_B) b_{Ba}(k_B)
\end{eqnarray}
and the counter terms are described by
\begin{eqnarray}
&& S_{CT} = \int_{k} \bar\psi_\sigma(k) \big( A_{\psi 1} i \zeta_\psi {\bm \Gamma} \cdot {\bm K} + A_{\psi 2} i\gamma_{d-1} k_{d-1} \big) \psi_\sigma(k) \nn
&&+ \int_k b_a^\dagger(k) \big( A_{b1} {\bm K}^2 + A_{b2} \vec k^2 \big) b_a(k) \nn &&
+ A_\lambda \frac{\lambda\mu^\epsilon}{4N} \int_{k,p,q} b_a^\dagger(k+q) b_a(k) b_b^\dagger(p-q) b_b(p) \nn
&& + \frac{1}{2} \int_q a(-q) \big( A_{a1} \zeta_a^2 {\bm Q}^2 + A_{a2} \vec q^2 \big) a(q) \nn &&
+ A_{\psi a} \frac{ie\mu^{\epsilon\over 2}}{\sqrt{N}} \int_{k,q} a(q) \bar\psi_\sigma(k+q) \gamma_5 \gamma_{d-1} \psi_\sigma(k) \nn
&&- A_{ba1} \frac{2e\mu^{\epsilon\over2}}{\sqrt{N}} \int_{k,q} \vec k \cdot \vec a(q) b_a^\dagger(k+q) b_a(k) \nn &&
+ A_{ba2} \frac{e^2 \mu^\epsilon}{N} \int_{k,p,q} \vec a(-p+q) \cdot \vec a(p) b_a^\dagger(k+q) b_a(k) .
\end{eqnarray}

Relations between bare and renormalized quantities are given by
\begin{eqnarray}
&&{\bm K} = \left( \frac{Z_{b2}}{Z_{b1}} \right)^{1\over2} {\bm K}_B, \ \ \
\vec k = \vec k_B, \nn
&&b(k) = Z_b^{-{1\over2}} b_B(k_B), \ \ \
Z_b = Z_{b2} \left( \frac{Z_{b2}}{Z_{b1}} \right)^{d-1\over2}, \nn
&&\psi_\sigma(k) = Z_{\psi}^{-{1\over2}} \psi_{B\sigma}(k_B), \ \ \
Z_\psi = Z_{\psi 2} \left( \frac{Z_{b2}}{Z_{b1}} \right)^{d-1\over2}, \nn
&& a(q) = Z_a^{-{1\over2}} a_B(q_B), \ \ \
Z_a = Z_{a2} \left( \frac{Z_{b2}}{Z_{b1}} \right)^{d-1\over2}, \nn
&& e_B^2 = e^2\mu^{\epsilon} Z_{a2}^{-1} \left( \frac{Z_{b2}}{Z_{b1}} \right)^{d-1\over2}, \nn
&&\lambda_B = \lambda \mu^\epsilon Z_\lambda Z_{b2}^{-2} \left( \frac{Z_{b2}}{Z_{b1}} \right)^{d-1\over2}, \nn
&&\zeta_{\psi B}^2 = \zeta_\psi^2 \left( \frac{Z_{\psi 1}}{Z_{\psi 2}} \right)^2 \frac{Z_{b2}}{Z_{b1}}, \ \ \
\zeta_{aB}^2 = \zeta_{a}^2 \frac{Z_{a1}}{Z_{a2}} \frac{Z_{b2}}{Z_{b1}}. \label{Scaling_Conformal_Invariance}
\end{eqnarray}

The spinon propagator is
\begin{eqnarray}
G_0^\psi(k) = -i \frac{\zeta_\psi {\bm \Gamma} \cdot {\bm K} + \gamma_{d-1} k_{d-1}}{\zeta_\psi^2 {\bm K}^2 + k_{d-1}^2},
\end{eqnarray}
the boson propagator is
\begin{eqnarray}
G_0^b(k) = \frac{1}{{\bm K}^2 + \vec k^2},
\end{eqnarray}
and the gauge-field propagator is
\begin{eqnarray}
G_0^a(q) = \frac{1}{\zeta_a^2 {\bm Q}^2 + \vec q^2}.
\end{eqnarray}
An essential point is that Landau damping does not occur in gauge fluctuations. The absence of Landau damping originates from the emergent Lorentz invariance. Such well propagating spin-singlet fluctuations cause much stronger effects on the Luttinger-liquid dynamics of spinons.

\subsubsection{Evaluation of counter terms in the one-loop level}

\begin{figure}[t]
\includegraphics[width=8cm]{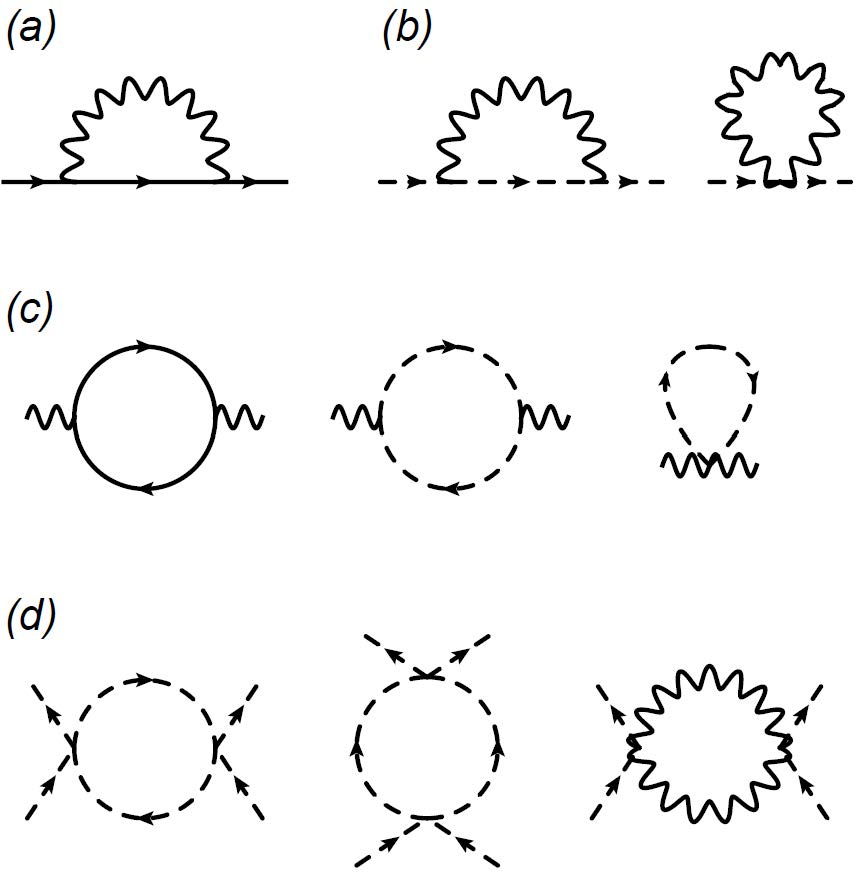}
\caption{Quantum corrections in the one-loop order. Self-energy corrections for spinons (a) and holons (b). Polarization bubbles from spinons and holons (c). $\lambda$-vertex corrections (d).}
\label{fig:SE_2}
\end{figure}

It is straightforward to evaluate quantum corrections in the one-loop level based on the dimensional regularization. First, we consider the role of U(1) gauge fluctuations in both the spinon and holon dynamics. The fermion self-energy given by the Fock diagram (Fig. \ref{fig:SE_2} (a)) is
\begin{eqnarray}
\Sigma^f(k)
&&= \left( \frac{ie\mu^{\epsilon\over2}}{\sqrt{N}} \right)^2 \int\frac{d^{d+1}q}{(2\pi)^{d+1}} \nn
&&\hspace{30pt} \times \gamma_5\gamma_{d-1} G_0^\psi(k+q) \gamma_5 \gamma_{d-1} G_0^a(q) \nn
&&= \frac{e^2}{8\pi^2N} \frac{1}{\epsilon} \Big[ -A i\zeta_\psi \bm \Gamma \cdot \bm K + Bi\gamma_{d-1} k_{d-1} \Big] + \mathcal{O}(\epsilon^0), \nn
\end{eqnarray}
where
\begin{eqnarray}
A &=& A(\zeta_\psi, \zeta_a)
= \int_0^1\!ds \ \frac{s^{1\over2}\zeta_a^2}{\left[ s\zeta_a^2 + (1-s) \zeta_\psi^2 \right]^2}, \nn
B &=& B(\zeta_\psi, \zeta_a)
= \int_0^1\!ds \ \frac{s^{1\over2}}{s\zeta_a^2 + (1-s) \zeta_\psi^2}. 
\end{eqnarray}
The boson self-energy given by the Fock diagram (Fig. \ref{fig:SE_2} (b)-left) is
\begin{eqnarray}
\Sigma^b (k)
&&= \left( -\frac{2e\mu^{\epsilon\over2}}{\sqrt{N}} \right)^2 \int\frac{d^{d+1}q}{(2\pi)^{d+1}} \nn
&&\hspace{50pt} \times (\vec k \cdot \hat t_q)^2 G_0^b(k+q) G_0^a(q) \nn
&&= \frac{e^2}{4\pi^2N\epsilon} \frac{\log \zeta_a^2}{\zeta_a^2 -1} \vec k^2 + \mathcal{O}(\epsilon^0),
\end{eqnarray}
where $\hat t_q$ is a unit vector perpendicular to $\vec q$. Recall that $\zeta_\psi$ and $\zeta_a$ describe anisotropic scaling between frequency and momentum in the spinon and gauge-field dispersions, respectively.

For the self-energy correction in U(1) gauge fluctuations, given by polarization functions, there are two contributions, the spinon bubble ($\Pi_\psi$) and the boson bubble ($\Pi_b$). The fermion polarization function (Fig. \ref{fig:SE_2} (c)-first) is given by
\begin{eqnarray}
\Pi_\psi(q)
&&= - 2\times \left( \frac{ie\mu^{\epsilon\over2}}{\sqrt{N}} \right)^2 \int\!\frac{d^{d+1}k}{(2\pi)^{d+1}} \nn
&&\times \mathrm{tr} \Big[G_0^\psi(k) \gamma_5\gamma_{d-1} G_0^\psi(k+q) \gamma_5 \gamma_{d-1} \Big] \nn
&&= \frac{e^2\mu^\epsilon}{N} \Lambda_d \frac{1}{12\zeta_\psi^2} \big( \zeta_\psi^2 {\bm Q}^2 + q_{d-1}^2 \big)^{1\over2} + \mathcal{O}(\epsilon).
\end{eqnarray}
Here, the minus sign comes from the spinon loop, and the factor $2$ is due to the $\sigma$-summation. $\mathrm{tr}$ denotes trace over Dirac gamma matrix space, and $\Lambda_d$ is a cut off in the $k_d$ direction. Note that there is no $1/\epsilon$-divergence. The holon polarization function (Fig. \ref{fig:SE_2} (c)-second) is given by
\begin{eqnarray}
\Pi_b(q)
&&= N\times \left( -\frac{2e\mu^{\epsilon\over2}}{\sqrt{N}} \right)^2 \int\frac{d^{d+1}k}{(2\pi)^{d+1}} \nn
&&\hspace{30pt} \times (\vec k \cdot \hat t_q)^2 G_0^b(k+q) G_0^b(k) \nn
&&= -\frac{e^2}{24\pi^2 \zeta_a^2\epsilon} \zeta_a^2 \bm Q^2 - \frac{e^2}{24\pi^2 \epsilon} \vec k^2 + \mathcal{O}(\epsilon^0) .
\end{eqnarray}

Second, we take into account the role of self-interactions in the holon dynamics. There are no quantum corrections in the self-energy of holons up to the one-loop level. Only vertex corrections (Fig. \ref{fig:SE_2} (d)) appear, given by
\bqa
\Gamma^{(1)}_\lambda(k,p;q)
&=& - \frac{N+5}{2} \left(\frac{\lambda \mu^{\epsilon}}{N}\right)^2 \int\!\frac{d^{d+1}k'}{(2\pi)^{d+1}} \nn &&\times G_0^b(k') G_0^b(k'+q) \nn
&=& -\frac{(N+5)\lambda^2}{16\pi^2N^2 \epsilon} + \mathcal{O}(\epsilon^0), \\
\Gamma^{(2)}_\lambda(k,p;q)
&=& - \left( \frac{e^2\mu^\epsilon}{N} \right)^2 \int\!\frac{d^{d+1}k'}{(2\pi)^{d+1}} \nn &&\times G_0^a(k') G_0^a(k'+q) \nn
&=& - \frac{e^4}{2\pi^2N^2 \zeta_a^2 \epsilon} + \mathcal{O}(\epsilon^0).
\eqa

Until now, we calculated the spinon self-energy correction from scattering with U(1) gauge fluctuations, the holon self-energy correction from scattering with U(1) gauge fluctuations, the gauge-field self-energy correction from scattering with both spinons and holons, and the vertex correction of the self-interaction term in the holon dynamics. Both interaction vertices between spinons and U(1) gauge fields and between holons and U(1) gauge fluctuations are determined straightforwardly, taking into account the Ward identity. As a result, we found the following counter terms
\begin{eqnarray}
&&A_{b1}= 0, \ \ \ A_{b2} = \frac{e^2}{4\pi^2N \epsilon} \frac{\log\zeta_a^2}{\zeta_a^2 -1}, \nn
&&A_{\psi 1} = -\frac{e^2}{8\pi^2N\epsilon} A(\zeta_\psi, \zeta_a), \ \ \
A_{\psi 2} = \frac{e^2}{8\pi^2N\epsilon} B(\zeta_\psi, \zeta_a), \nn
&&A_{a1} = -\frac{e^2}{24\pi^2 \zeta_a^2 \epsilon}, \ \ \
A_{a2} =  -\frac{e^2}{24\pi^2\epsilon}, \nn
&&A_\lambda = \frac{(N+5)\lambda}{16\pi^2N\epsilon} + \frac{e^4}{2\pi^2N\lambda \zeta_a^2 \epsilon}, \nn &&
A_{\psi a} = \frac{e^2}{8\pi^2N\epsilon} B(\zeta_\psi, \zeta_a), \ \ \
A_{ba} = \frac{e^2}{4\pi^2N \epsilon} \frac{\log\zeta_a^2}{\zeta_a^2 -1} . \label{Counter_Terms_Conformal_Invariance}
\end{eqnarray}
We note $A_{\psi 2} = A_{\psi a}$ and $A_{b2} =  A_{ba1} = A_{ba2} \equiv A_{ba}$, which result from the Ward identity.

\subsubsection{Renormalization group equations}

It is straightforward to express all the scaling equations of Eq. (\ref{Scaling_Conformal_Invariance}) in the form of differential equations, given by the fact that all bare quantities do not change under the scaling transformation varying the mass scale of $\mu$. As a result, we obtain general expressions for flow equations of interaction parameters and dispersion coefficients as follows:
\begin{eqnarray}
\frac{\mu}{e^2} \frac{de^2}{d\mu}
&=& -\epsilon + \frac{\mu}{Z_{a2}} \frac{dZ_{a2}}{d\mu} - \frac{2-\epsilon}{2} \frac{\mu}{Z_{b2}/Z_{b1}} \frac{d(Z_{b2}/Z_{b1})}{d\mu}, \nn
\frac{\mu}{\lambda} \frac{d\lambda}{d\mu}
&=& -\epsilon - \frac{\mu}{Z_\lambda} \frac{dZ_\lambda}{d\mu} + 2\frac{\mu}{Z_{b2}} \frac{dZ_{b2}}{d\mu}\nn
&&- \frac{2-\epsilon}{2} \frac{\mu}{Z_{b2}/Z_{b1}} \frac{d(Z_{b2}/Z_{b1})}{d\mu} , \nn
\frac{\mu}{\zeta_\psi^2} \frac{d\zeta_\psi^2}{d\mu}
&=& - \frac{\mu}{Z_{b2}/Z_{b1}} \frac{d(Z_{b2}/Z_{b1})}{d\mu} + 2\frac{\mu}{Z_{\psi 2}/Z_{\psi 1}} \frac{d(Z_{\psi 2}/Z_{\psi 1})}{d\mu} \nn
\frac{\mu}{\zeta_a^2} \frac{d\zeta_a^2}{d\mu}
&=& - \frac{\mu}{Z_{b2}/Z_{b1}} \frac{d(Z_{b2}/Z_{b1})}{d\mu} + \frac{\mu}{Z_{a 2}/Z_{a 1}} \frac{d(Z_{a 2}/Z_{a 1})}{d\mu} , \nn
\end{eqnarray}
which show how such parameters scale as a function of $\mu$. Inserting renormalization factors given by counter terms of Eq. (\ref{Counter_Terms_Conformal_Invariance}) into the above, we find $\beta-$functions in two dimensions ($\epsilon=1$)
\begin{eqnarray}
\beta_e
&\equiv& \mu\frac{de^2}{d\mu}
= e^2 \left( - 1 + \frac{e^2}{24\pi^2} + \frac{e^2}{8\pi^2 N} \frac{\log \zeta_a^2}{\zeta_a^2-1} \right), \nn
\beta_\lambda
&\equiv& \mu\frac{d\lambda}{d\mu}
= \lambda \bigg( - 1 + \frac{(N+5)\lambda}{16\pi^2N} + \frac{e^4}{2\pi^2 N \lambda \zeta_a^2}  \nn && - \frac{3e^2}{8\pi^2N} \frac{\log \zeta_a^2}{\zeta_a^2 -1} \bigg), \nn
\beta_{\zeta_\psi}
&\equiv& \mu\frac{d\zeta_\psi^2}{d\mu}
= \zeta_\psi^2 \bigg( \frac{e^2}{4\pi^2N} \frac{\log \zeta_a^2}{\zeta_a^2 -1} \nn
&&\hspace{45pt} - \frac{e^2}{4\pi^2 N} \Big[ A(\zeta_\psi, \zeta_a) + B(\zeta_\psi, \zeta_a) \Big] \bigg), \nn
\beta_{\zeta_a}
&\equiv& \mu\frac{d\zeta_a^2}{d\mu}
= \zeta_a^2 \bigg( \frac{e^2}{4\pi^2N} \frac{\log \zeta_a^2}{\zeta_a^2 -1} + \frac{e^2}{24\pi^2} - \frac{e^2}{24\pi^2\zeta_a^2} \bigg). \nn
\label{RG_Equations_Conformal_Symmetry}
\end{eqnarray}

\begin{figure}[t]
\includegraphics[width=9cm]{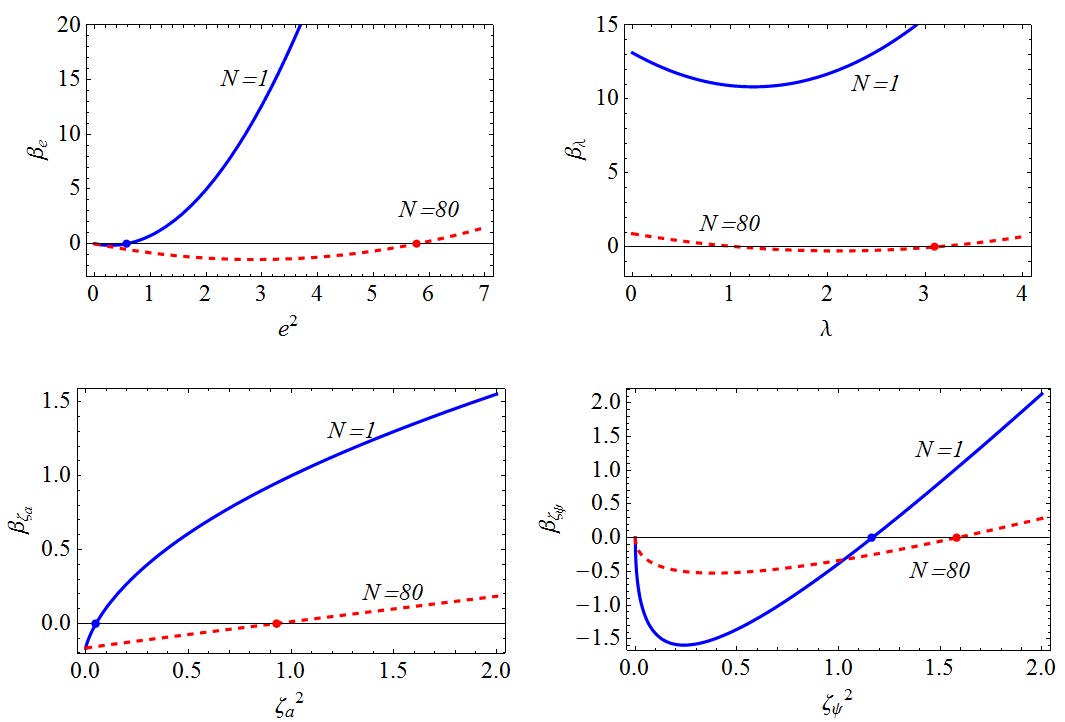}
\caption{Renormalization group flows from Eq. (\ref{RG_Equations_Conformal_Symmetry}) for spin-liquid Mott quantum criticality. }
\label{fig:Fixed_Points}
\end{figure}

Renormalization group flows are summarized in Fig. \ref{fig:Fixed_Points}. Figure \ref{fig:Fixed_Points} (a) shows the evolution of the gauge charge as a function of the scaling parameter. The tree-level scaling analysis shows the relevance of the gauge coupling, regarded to be a trivial result below the upper critical dimension. On the other hand, such gauge fluctuations should be screened by quantum corrections given by polarizations of spinons and holons, resulting in a finite critical value of the interaction parameter. Figure \ref{fig:Fixed_Points} (b) shows the flow of the self-interaction parameter $\lambda$. The evolution of the self-interaction parameter $\lambda$ as a function of the scaling parameter $\mu$ is consistent with many previous results \cite{IXY_Review,Fluctuation_Driven_First_Order_I,Fluctuation_Driven_First_Order_II}. When the flavor number $N$ of holons is smaller than a critical value, here $N_{c} = 58$, the $\beta_{\lambda}$ function is always positive, which shows that the self-interaction constant flows into a negative value, implying the first-order condensation transition of holons. This is well known to be either the Coleman-Weinberg mechanism in high energy physics \cite{Fluctuation_Driven_First_Order_II} or the fluctuation-driven first-order phase transition in condensed matter physics \cite{Fluctuation_Driven_First_Order_I}. On the other hand, when the holon flavor number is larger than the critical value, the $\beta_{\lambda}$ function allows a stable critical fixed point, given by the second zero between $3$ and $4$ in the case of $N = 80$. The first zero point around $1$, regarded to be an unstable fixed point, is suggested to be a tri-critical point, which distinguishes the first order transition from the second order one \cite{IXY_Review}. Figure \ref{fig:Fixed_Points} (c) shows the change of the anisotropy between frequency and momentum, or inverse of velocity of holons. This flows to a stable fixed point, given by $\zeta_{a} = \zeta_{a*}$. The renormalization group flow for the scaling anisotropy between frequency and momentum in the free part of the spinon field, Fig. \ref{fig:Fixed_Points} (d), also gives rise to a stable fixed point of $\zeta_\psi = \zeta_{\psi*}$ in the IR limit. The scaling anisotropy between frequency and momentum in the free part of the U(1) gauge field and the spinon field turns out to be not important at low energies, regarded to still show the relativistic invariance approximately.

In order to clarify the existence of the second-order phase transition, we take into account the $N\rightarrow \infty$ limit, where the beta functions become
\begin{eqnarray}
&&\beta_e = e^2 \left( -1 + \frac{e^2}{24\pi^2} \right), \ \ \
\beta_\lambda = \lambda \left( -1 + \frac{\lambda}{16\pi^2} \right), \nn
&&\beta_{\zeta_\psi} = 0, \ \ \
\beta_{\zeta_a} = \zeta_a^2 \left( \frac{1}{6} - \frac{1}{6\zeta_a^2} \right) ,
\end{eqnarray}
which result in a fixed point, given by $e_*^2/(4\pi^2) = 6$, $\lambda_*/(4\pi^2) = 4$, and $\zeta_{a*}^2 = 1$. Here, $\zeta_\psi^2$ is marginal. The fixed point values are summarized in Table \ref{table:FPs}.

\begin{table}[b]
\caption{\label{tab:table1} Fixed point values, critical exponents, and anomalous scaling dimensions.}
\begin{ruledtabular}
\begin{tabular}{cccc}
 & $N=1$ & $N=80$ & $N\rightarrow \infty$ \\
\colrule
$e_*^2/(4\pi^2)$     & 0.5744  & 5.775    & 6 \\
$\lambda_*/(4\pi^2)$ & -       & 3.094    & 4 \\
$\zeta_a^2$          & 0.05027 & 0.9278   & 1 \\
$\zeta_\psi^2$       & 1.164   & 1.578    & 1.601 \\
\colrule
$z_*$                & 1.904   & 1.004    & 1 \\
$\eta_{\psi*}$       & -0.7882 & -0.02907 & 0 \\
$\eta_{b*}$          & -1.356  & -0.05620 & 0 \\
$\eta_{a*}$          & -0.4043 & 0.4625   & 1/2 \\
\end{tabular}
\end{ruledtabular}
\label{table:FPs}
\end{table} 

\subsubsection{Callan-Symanzik equation}

In order to understand physical properties near the spin-liquid Mott critical point, we should find the scaling theory for correlation functions. The scaling theory is given by the solution of the Callan-Symanzik equation, a differential equation for correlation functions to satisfy, describing the evolution of correlation functions as a function of the scaling parameter $\mu$ \cite{Callan_Symanzik_Eqs}.

The Callan-Symanzik equation for our model is (Appendix \ref{App:CS_Eq})
\begin{eqnarray}
&&\bigg[ z {\bm K}_i \cdot \nabla_{{\bm K}_i} + \vec k_i \cdot \nabla_{\vec k_i} \nn && - \beta_e \frac{\partial}{\partial e^2} - \beta_\lambda \frac{\partial}{\partial \lambda} - \beta_{\zeta_\psi} \frac{\partial}{\partial \zeta_\psi} - \beta_{\zeta_a} \frac{\partial}{\partial \zeta_a} \nn && - 2m \left( - \frac{5-\epsilon}{2} + \eta_\psi \right) - 2n \left( - \frac{6-\epsilon}{2} + \eta_b \right) \nn && - 2l \left( - \frac{6-\epsilon}{2} + \eta_a \right) - \{ z(2-\epsilon) + 2 \} \bigg] \nn && G^{(m,n,l)} \Big( \{ k_i \}; e, \lambda, \zeta_\psi, \zeta_a, \mu \Big) = 0,
\label{eq:CS_eq}
\end{eqnarray}
where $G^{(m,n,l)}$ is renormalized $(m+n+l)$-point Green's function given by
\begin{eqnarray}
&&\Big< \bar\psi(k_1) \cdots \bar\psi(k_m) \psi(k_{m+1}) \cdots \psi(k_{2m}) \nn
&&\times b^\dagger(k_{2m+1}) \cdots b^\dagger(k_{2m+n}) b(k_{2m+n+1}) \cdots b(k_{2m+2n}) \nn
&&\ \times a(k_{2m+2n+1}) \cdots a(k_{2m+2n+2l}) \Big> \nn
&&= G^{(m,n,l)} \Big( \{ k_i \}; e, \lambda, \zeta_\psi, \zeta_a, \mu \Big) \ \delta^{(d+1)} \Big( \{ k_i \} \Big).
\end{eqnarray}
The beta functions $\beta_g$, $g=e, \lambda, \zeta_\psi, \zeta_a$ are defined in Eq. (\ref{RG_Equations_Conformal_Symmetry}). The dynamical critical exponent $z$ and anomalous scaling dimensions $\eta_i$, $i=\psi, b, a$ are given by
\begin{eqnarray}
&&z = 1- \frac{1}{2} \frac{\mu}{Z_{b2}/Z_{b1}} \frac{\partial Z_{b2}/Z_{b1}}{\partial \mu}, \nn
&&\eta_\psi = \frac{1}{2} \frac{\mu}{Z_\psi} \frac{\partial Z_\psi}{\partial \mu}, \ \ \
\eta_b = \frac{1}{2} \frac{\mu}{Z_b} \frac{\partial Z_b}{\partial \mu}, \ \ \
\eta_a = \frac{1}{2} \frac{\mu}{Z_a} \frac{\partial Z_a}{\partial \mu}. \nn
\end{eqnarray}
These values are evaluated in the one-loop level at the fixed point, summarized in Table \ref{table:FPs}.

%
%

Solving the Callan-Symanzik equation at the fixed point, we obtain the spinon Green's function
\begin{eqnarray}
G_{\psi \sigma}({\bm K}, k_{d-1})
&=& \left< \bar\psi_\sigma(k) \psi_\sigma(k) \right> \nn
&=& \frac{1}{|k_{d-1}|^{2-z_*-2\eta_{\psi *}}} f_{\psi\sigma} \left( \frac{|{\bm K}|^{1/z}}{|k_{d-1}|} \right), \nn
\end{eqnarray}
where $f_{\psi\sigma}$ is a non-singular function, which can be determined by a direct calculation. Here, no $\sigma$-summation is performed. The exponent of $|k_{d-1}|$ is
\begin{eqnarray}
2-z_*-2\eta_{\psi *} \approx
\left\{ \begin{array}{ll}
1.672  & (N=1), \\
1.054  & (N=80), \\
1      & (N\rightarrow \infty),
\end{array}\right.
\end{eqnarray}
positive regardless of $N$.

\subsection{Renormalization of a $q = 2 k_{F}$ vertex}

In order to clarify effects of the disappearance of a spinon Fermi surface on physical responses, we consider a renormalization group flow for a $2 k_{F}$ vertex with its strength of $r$, given by 
\begin{eqnarray}
&& S_r = - 2 r \mu \int\!\frac{d^3k}{(2\pi)^3} \nn && \hspace{40pt} \left( f_{\sigma +}^\dagger(k) f_{\sigma -}(k) + f_{\sigma -}^\dagger(k) f_{\sigma +}(k) \right) \nn
&& \rightarrow i r \mu \int\!\frac{d^{d+1}k}{(2\pi)^{d+1}} \nn && \hspace{40pt} \Big( \psi^T(k) \gamma_0 \psi(-k) + \bar\psi(k) \gamma_0 \bar\psi^T(-k) \Big) ,
\end{eqnarray}
where the dimensional regularization has been introduced in the last line. Then, the renormalization of this vertex is given by
\begin{eqnarray}
&& \Gamma_r = r \left( \frac{ie\mu^{\epsilon\over2}}{\sqrt{N}} \right)^2 \int\!\frac{d^{d+1}q}{(2\pi)^{d+1}} \ G_0^a(q) \big( \gamma_5\gamma_{d-1} \big)^T \nn && \big[G_0^\psi\big]^T(k+q) \gamma_0 G_0^\psi(-k-q) \gamma_5\gamma_{d-1}
\end{eqnarray}
in the one-loop level.

\begin{figure}[t]
\includegraphics[width=8cm]{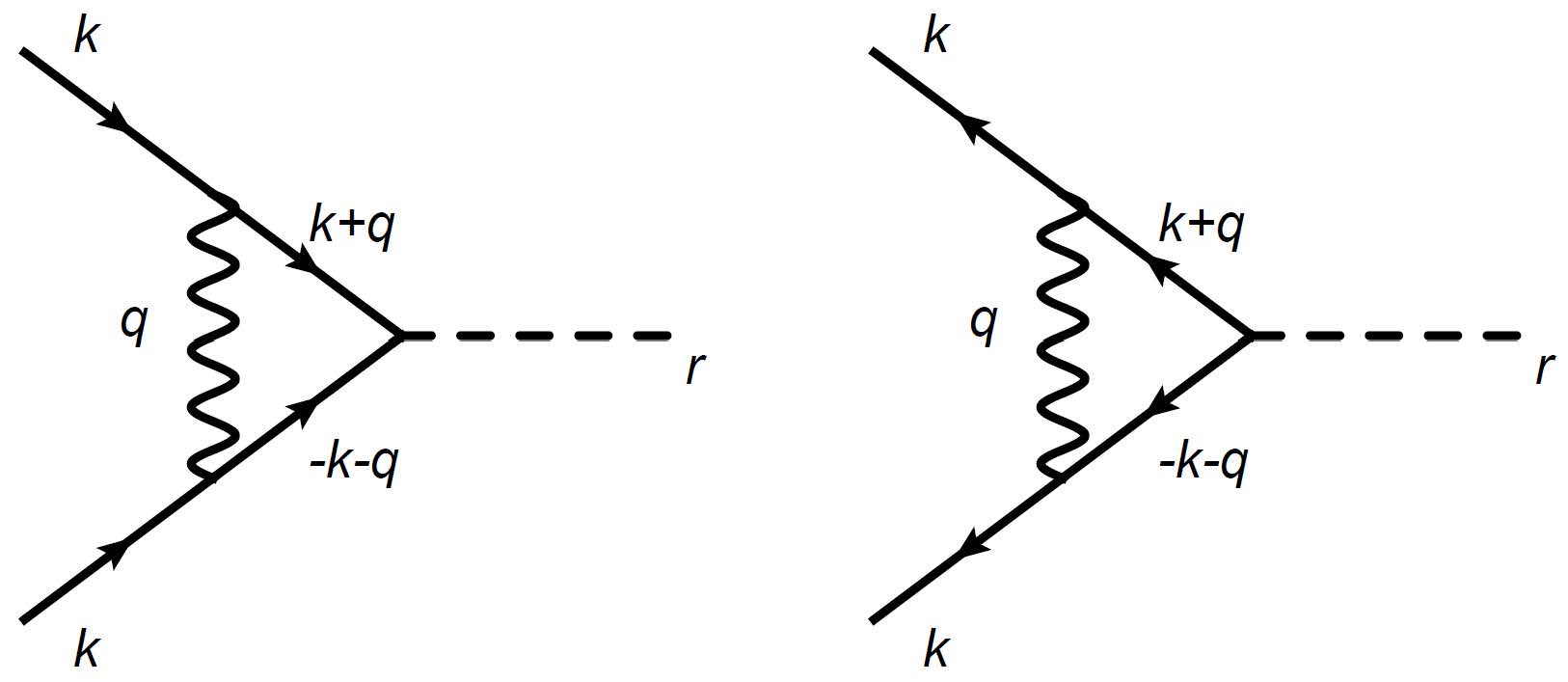}
\caption{Renormalization of a $2 k_{F}$ vertex with its strength of $r$.}
\label{fig:2kF_vertex}
\end{figure}

When the U(1) spin-liquid state with a spinon Fermi surface is considered, the $\beta_r(\mu)$ function is given by \cite{SSL_Dimensional_Regularization_Nematic} 
\begin{eqnarray}
\beta_r(\mu)
&\equiv& \mu\frac{dr}{d\mu} = r \left[ -1 - \frac{\mu}{Z_r} \frac{dZ_r}{d\mu} + \frac{\mu}{Z_{\psi 2}}\frac{Z_{\psi2}}{d\mu} \right] \nn
&\approx& 0.0772 r ,
\end{eqnarray} 
where the vertex renormalization constant is 
\begin{eqnarray}
Z_r = 1 + u_1 \frac{e^{4\over3}}{\epsilon}, \ \ \ u_1 \approx 0.1346 .
\end{eqnarray}
This shows irrelevance of the $2 k_{F}$ scattering channel in the presence of a spinon Fermi surface.

On the other hand, if the spin-liquid Mott quantum critical point is taken into account, the $\beta_r(\mu)$ function is given by 
\begin{eqnarray}
\beta_r(\mu)
&\equiv& \mu\frac{dr}{d\mu} = r \left[ -1 - \frac{\mu}{Z_r} \frac{dZ_r}{d\mu} + \frac{\mu}{Z_{b 2}}\frac{Z_{b2}}{d\mu} \right] \nn
&\approx& \left\{ \begin{array}{ll}
- 1.440 r  & (N=1), \\
- 1.004 r  & (N=80), \\
- 1.000 r  & (N \rightarrow \infty),
\end{array}\right. 
\end{eqnarray} 
respectively. Here, the vertex renormalization constant is  
\begin{eqnarray}
&& Z_r = 1 + \frac{e^2}{16\pi^2N\epsilon} C_r(\zeta_\psi, \zeta_a), \nn
&& C_r(\zeta_\psi, \zeta_a) = \int_0^1\!ds \ \frac{1-s}{s^{1\over2} \left[ s\zeta_a^2 + (1-s)\zeta_\psi^2 \right]}.
\end{eqnarray} 
As a result, one-dimensional spinon dynamics gives rise to the enhancement of spin correlations for the $2 k_{F}$ channel.

\section{Renormalization group analysis III: Bosonization for spinons}

\subsection{Bosonization for spinons}

The curvature term with $k_2^2$ in the spinon spectrum is irrelevant in the scaling analysis, being set to be zero. In other words, the dispersionless dispersion along the $k_2$-direction tells that the spinon dynamics is localized in the $x_2$-direction. As a result, we start from the following effective field theory in two dimensions
\begin{eqnarray}
S
&=& \int\!d^2x \ \bar\Psi_\sigma (k) \big( \gamma_0 \partial_0 + \gamma_1 \partial_1 \big) \Psi_\sigma(k) \nn
&+& ie \int\!d^2x  \  A(x) \bar\Psi_\sigma(x)\gamma_5 \gamma_1 \Psi_\sigma(x) \nn
&+& \int\!d^3x \ \big| \big( \partial_\mu - iea_\mu(x) \big) b(x) \big|^2 + \frac{\lambda}{4} \int\!d^3x \ \big| b(x) \big|^4 \nn
&+& \frac{1}{2} \int\!d^3x \ \big( \partial_\mu a(x) \big)^2.
\end{eqnarray}
Here, $\psi_\sigma(x_0, x_1, x_2) = \Psi_\sigma(x_0, x_1) \delta(x_2)$ and $a(x_0, x_1, x_2 = 0) = A(x_0, x_1)$ are effective one dimensional spinons and U(1) gauge fields, which emerge at low energies near the spin-liquid Mott quantum critical point. We recall $\mu = 0, 1, 2$ with $a_{0} = 0$ in the holon sector.

Since the critical dynamics of spinons are effectively described by Luttinger-liquid physics, we can use the bosonization technique \cite{Luttinger_Liquid_Review} to solve the spinon sector. Performing the bosonization for $\Psi_\sigma$, we obtain
\begin{eqnarray}
S &=& \frac{1}{2\pi} \int\!d^2x \ \bigg[ \big( \partial_\mu \Phi_c(x) \big)^2 + \big( \partial_\mu \Phi_s(x) \big)^2 \nn &-& 2ieA(x) \partial_1 \Phi_c(x) \bigg]
+ \int\!d^3x \ \big| \big( \partial_\mu - iea_\mu(x) \big) b(x) \big|^2 \nn &+& \frac{\lambda}{4} \int\!d^3x \ \big| b(x) \big|^4 + \frac{1}{2} \int\!d^3x \ \big( \partial_\mu a(x) \big)^2.
\end{eqnarray}
Here, $\Phi_{c}(x)$ represents neutral density fluctuations (sound modes) and $\Phi_{s}(x)$ describes collective spin density excitations of the Ising type. U(1) gauge fluctuations couple to neutral density excitations as expected. Integrating over $\Phi_c(x)$, we obtain
\begin{eqnarray}
S
&=& \frac{1}{2\pi} \int\!d^2x \ \big( \partial_\mu \Phi_s(x) \big)^2 \nn
&+& \int\!d^3x \ \big| \big( \partial_\mu - iea_\mu(x) \big) b(x) \big|^2 + \frac{\lambda}{4} \int\!d^3x \ \big| b(x) \big|^4 \nn
&+& \frac{1}{2} \int\!d^3x \ \big( \partial_\mu a(x) \big)^2 + \frac{e^2}{2\pi} \int\!d^2x \ A^2(x).
\end{eqnarray}
Now, the $\Phi_s$ field is decoupled to the rest of the fields. The critical spin dynamics is described by the Luttinger liquid theory. Recall that we performed the Abelian bosonization. If we resort to the non-Abelian bosonization technique, respecting the spin SU(2) symmetry, we obtain SU(2) $k = 1$ Wess-Zumino-Witten (WZW) theory \cite{Luttinger_Liquid_Review} for the critical spin dynamics near the spin-liquid Mott quantum critical point, where $k$ denotes the level of the theory. On the other hand, U(1) gauge fluctuations look ``massive". However, one should be more careful to reach such a conclusion since the $A(x)$ field exists only at $x_{2} = 0$.

In order to take into account the delta function mass-like term, it is convenient to use the following scattering basis for the gauge field $a$:
\begin{eqnarray}
u_+(x_2;k_2) &=&
\left\{
\begin{array}{ll}
e^{ik_2x_2} - \frac{i\alpha_k}{1+i\alpha_k} e^{-ik_2x_2} & (x_2<0), \\ \\
\frac{1}{1+i\alpha_k} e^{ik_2x_2} & (x_2>0),
\end{array}
\right. \\ \nn \nn
u_-(x_2;k_2) &=&
\left\{
\begin{array}{ll}
\frac{1}{1+i\alpha_k} e^{-ik_2x_2} & (x_2<0), \\ \\
e^{-ik_2x_2} - \frac{i\alpha_k}{1+i\alpha_k} e^{ik_2x_2} & (x_2>0),
\end{array}
\right.
\end{eqnarray}
where $u_\pm$ orthonormal;
\begin{eqnarray}
&&\int\!dx_2 \ u_+^*(x_2;k_2') u_+(x_2;k_2) = (2\pi) \delta(k_2-k_2'), \nn
&&\int\!dx_2 \ u_-^*(x_2;k_2') u_-(x_2;k_2) = (2\pi) \delta(k_2-k_2'), \nn
&&\int\!dx_2 \ u_-^*(x_2;k_2') u_+(x_2;k_2) = 0.
\end{eqnarray}

Expanding $a(x)$ in terms of these eigenfunctions, we obtain
\begin{eqnarray}
a(x) = \int\!\frac{d^3q}{(2\pi)^3} \ \Big[ a_+(q) u_+(x;q) + a_-(q) u_-(x;q) \Big].
\end{eqnarray}
As a result, we reach the following expression (extending the number of component of $b$-field to $N$)
\begin{eqnarray}
&& S = \frac{1}{2\pi} \int\!d^2x \ \big( \partial_\mu \Phi_s(x) \big)^2 \nn && + \int_k \ \big( k_0^2 + k_1^2 + k_2^2 \big) b_a^\dagger(k) b_a(k) \nn && - \frac{e}{2\sqrt{N}} \int_{k,q} \ (2k_\mu + q_\mu)
\Big[ a^+_\mu(q) + a^-_\mu(q) \Big] b_a^\dagger(k+q) b_a(k) \nn && + \frac{e^2}{2N} \int_{k,p,q} \ \Big[ a_+(-p+q) a_+(p) + a_-(-p+q) a_-(p) \Big] \nn
&&\times b_a^\dagger(k+q) b_a(k) \nn
&& + \frac{\lambda}{4N} \int_{k,p,q} \ b_a^\dagger(k+q) b_a(k) b_b^\dagger(p-q) b_b(p) \nn
&& + \frac{1}{2} \int_q \ \big( q_0^2 + q_1^2 + q_2^2 \big) \Big[ a_+(-q) a_+(q) + a_-(-q) a_-(q) \Big] , \nn
\end{eqnarray}
which remains essentially the same as the Abelian Higgs model. We conclude that U(1) gauge fluctuations are not massive for the role in critical holon dynamics.

\subsection{Renormalization group analysis for the holon sector}

It is straightforward to perform the renormalization group analysis for the holon sector based on the dimensional regularization technique. Taking into account the scaling transformation \bqa && k = \frac{k'}{s} , \eqa we obtain
\begin{eqnarray}
&&b(k) = s^{d+3\over2}b'(k'), \ \ \
a_\pm(q) = s^{d+3\over2} a_\pm'(q')
\end{eqnarray}
for field variables and
\begin{eqnarray}
&& e' = s^{3-d\over2} e, \ \ \
\lambda' = s^{3-d} \lambda.
\end{eqnarray}
for interaction parameters. Both coupling parameters are marginal at $d_{c} = 3$, and the renormalization group analysis is performed in $d = 3 - \epsilon$.

The $(d+1)$-dimensional effective field theory (bare action)
%
%
%
%
\begin{eqnarray}
&& S_B
= \int_{k_B} \ \big( {\bm K}_{B}^2 + \vec k_B^2 \big) b_{Ba}^\dagger(k_B) b_{Ba}(k_B) \nn
&& - \frac{e_B}{\sqrt{N}} \int_{k_B, q_B} \ \vec k_B \cdot \Big[ \vec a_{+,B}(q_B) + \vec a_{-,B}(q_B) \Big] \nn
&&\times b_{Ba}^\dagger(k_B+q_B) b_{Ba}(k_B) \nn
&& + \frac{e_B^2}{2N} \int_{k_B, p_B, q_B} \ \Big[ a_{+,B}(-p_B+q_B) a_{+,B}(p_B) \nn && + a_{-,B}(-p_B+q_B) a_{-,B}(p_B) \Big] b_{Ba}^\dagger(k_B+q_B) b_{Ba}(k_B) \nn && + \frac{\lambda_B}{4N} \int_{k_B, p_B, q_B} \ b_{Ba}^\dagger(k_B+q_B) b_{Ba}(k_B) \nn && b_{Bb}^\dagger(p_B-q_B) b_{Bb}(p_B) \nn
&& + \frac{1}{2} \int_{q_B} \ \big( \zeta_{aB}^2 {\bm Q}_B^2 + \vec q_B^2 \big) \Big[ a_{+,B}(-q_B) a_{+,B}(q_B) \nn && + a_{-,B}(-q_B) a_{-,B}(q_B) \Big]
\end{eqnarray}
is separated into the renormalized action
\begin{eqnarray}
&& S_R = \int_k \ \big( {\bm K}^2 + \vec k^2 \big) b_a^\dagger(k) b_a(k) \nn
&& - \frac{e\mu^{\epsilon\over2}}{\sqrt{N}} \int_{k,q} \ \vec k \cdot \Big[ \vec a_{+}(q) + \vec a_{-}(q) \Big] b_a^\dagger(k+q) b_a(k) \nn
&& + \frac{e^2\mu^\epsilon}{2N} \int_{k,p,q} \ \Big[ a_{+}(-p+q) a_{+}(p) + a_{-}(-p+q) a_{-}(p) \Big] \nn
&&\times b_a^\dagger(k+q) b_a(k) \nn
&& + \frac{\lambda\mu^\epsilon}{4N} \int_{k,p,q} \ b_a^\dagger(k+q) b_a(k) b_b^\dagger(p-q) b_b(p) \nn
&& + \frac{1}{2} \int_q \ \big( \zeta_a^2 {\bm Q}^2 + \vec q^2 \big) \Big[ a_{+}(-q) a_{+}(q) + a_{-}(-q) a_{-}(q) \Big] \nn
\end{eqnarray}
and counter terms
\begin{eqnarray}
&& S_{CT}
= \int_k \ \big( A_{b1}{\bm K}^2 + A_{b2}\vec k^2 \big) b_a^\dagger(k) b_a(k) \nn
&& - A_{ba1} \frac{e\mu^{\epsilon\over2}}{} \int_{k,q} \ (\vec k \cdot \hat t_q) \Big[ a_{+}(q) + a_{-}(q) \Big] b^\dagger(k+q) b(k) \nn
&& + A_{ba2} \frac{e^2\mu^\epsilon}{2} \int_{k,p,q} \ \Big[ a_{+}(-p+q) a_{+}(p) + a_{-}(-p+q) a_{-}(p) \Big] \nn
&&\times b^\dagger(k+q) b(k) \nn
&& + A_\lambda \frac{\lambda \mu^\epsilon}{4} \int_{k,p,q} \ b^\dagger(k+q) b^\dagger(p-q) b(p) b(k) \nn
&& + \frac{1}{2} \int_q \ \big( A_{a1} \zeta_a^2 {\bm Q}^2 + A_{a2} \vec q^2 \big) \Big[ a_{+}(-q) a_{+}(q) \nn
&& + a_{-}(-q) a_{-}(q) \Big] ,
\end{eqnarray}
where $\epsilon = 3-d$ and introduced mass scaling dimension $\mu$. The Ward identity guarantees $A_{b2} = A_{ba1} = A_{ba2}$.

The relation between bare and renormalized quantities are given by
\begin{eqnarray}
&& {\bm K} = \left( \frac{Z_{b2}}{Z_{b1}} \right)^{1\over2} {\bm K}_B, \ \ \
\vec k = \vec k_B, \nn
&& b_a(k) = Z_b^{-{1\over2}} b_{Ba}(k_B), \ \ \ Z_b = Z_{b2} \left( \frac{Z_{b2}}{Z_{b1}} \right)^{d-1\over2}, \nn
&&a_\pm(q) = Z_a^{-{1\over2}} a_{\pm, B}(q_B), \ \ \ Z_a = Z_{a2} \left( \frac{Z_{b2}}{Z_{b1}} \right)^{d-1\over2}, \nn
&&e_B^2 = e^2 \mu^{\epsilon} Z_{a2}^{-1} \left( \frac{Z_{b2}}{Z_{b1}} \right)^{d-1\over2}, \nn
&&\lambda_B = \lambda \mu^\epsilon Z_\lambda Z_{b2}^{-2} \left( \frac{Z_{b2}}{Z_{b1}} \right)^{d-1\over2}, \nn
&&\zeta_{aB}^2 = \zeta_a^2 \frac{Z_{b2}}{Z_{b1}} \frac{Z_{a1}}{Z_{a2}}.
\end{eqnarray}
\\

\subsection{Evaluation of counter terms in the one-loop level}

The self-energy correction of the gauge field is given by the polarization function of the holon field, given by
\begin{eqnarray}
\Pi_{a_\pm}(q)
&=& N\times \left( -\frac{e\mu^{\epsilon\over2}}{\sqrt{N}} \right)^2 \int\!\frac{d^{d+1}k}{(2\pi)^{d+1}} \nn && (\vec k \cdot \hat t_q)^2 G_0^b(k) G_0^b(k+q) \nn
&=& -\frac{e^2}{96\pi^2 \epsilon} ({\bm Q}^2 + \vec q^2) + \mathcal{O}(\epsilon^0) ,
\end{eqnarray}
where the holon propagator is
\begin{eqnarray}
G_0^b(k) = \frac{1}{{\bm K}^2 + \vec k^2} .
\end{eqnarray}
The holon self-energy correction is described by the Fock diagram, given by
\begin{eqnarray}
\Sigma_b(k)
&=& 2\left(-\frac{e\mu^{\epsilon\over2}}{\sqrt{N}} \right)^2 \int\!\frac{d^{d+1}q}{(2\pi)^{d+1}} \nn && (\vec k \cdot \hat t_q)^2 G_0^a(q) G_0^b(k+q) \nn
&=& \frac{e^2}{8\pi^2 N\epsilon} \frac{\log\zeta_a^2}{\zeta_a^2-1} \vec k^2 + \mathcal{O}(\epsilon^0) ,
\end{eqnarray}
where the gauge-field propagator is
\begin{eqnarray}
G_a(q) = \frac{1}{\zeta_a^2 {\bm Q}^2 + \vec q^2},
\end{eqnarray}
the factor $2$ comes from summation of two kinds of gauge fields.

The vertex correction in the holon-gauge vertex can be found, resorting to the Ward identity. The renormalization effect for the holon self-interaction vertex is well known to follow a textbook level. As a result, we obtain counter terms as follows
\begin{eqnarray}
&&A_{b1} = 0, \ \ \
A_{b2} = \frac{e^2}{8\pi^2 N \epsilon} \frac{\log \zeta_a^2}{\zeta_a^2 - 1}, \nn
&&A_{a1} = - \frac{e^2}{96\pi^2 \zeta_a^2 \epsilon}, \ \ \
A_{a2} = - \frac{e^2}{96\pi^2 \epsilon}, \nn
&&A_\lambda = \frac{(N+5) \lambda}{16\pi^2 N\epsilon} + \frac{e^4}{16\pi^2 N \lambda \zeta_a^2 \epsilon}, \nn
&&A_{ba1}= A_{ba2} = \frac{e^2}{8\pi^2 N \epsilon} \frac{\log \zeta_a^2}{\zeta_a^2 - 1}.
\end{eqnarray}
%
%

\subsection{Renormalization group equations}

\begin{figure*}[t]
\includegraphics[width=16cm]{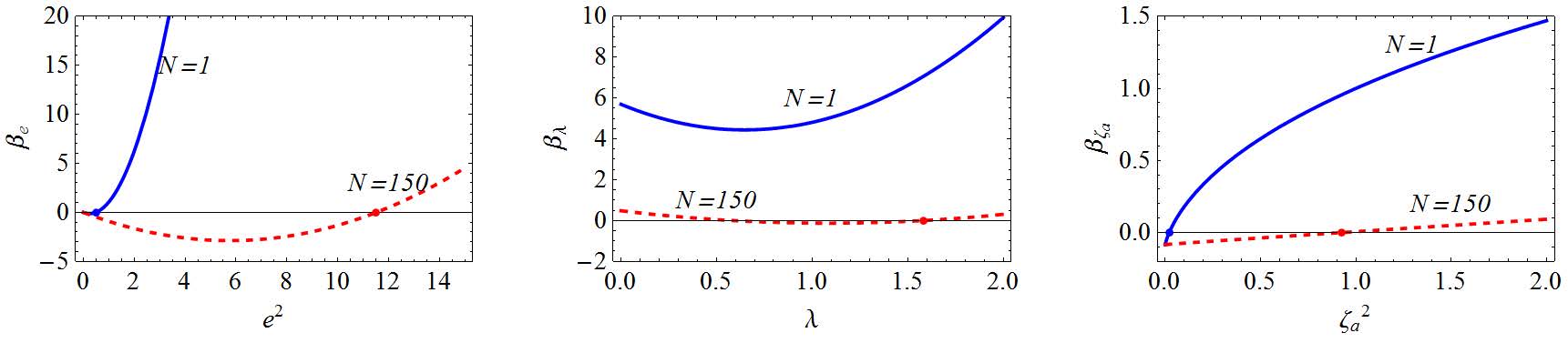}
\caption{Renormalization group flows from Eq. (\ref{RG_Bosonization}). They are essentially the same as those in Fig. \ref{fig:Fixed_Points}.}
\label{fig:FP_bosonization}
\end{figure*}

Considering that bare quantities do not evolve with respect to the scaling parameter $\mu$, it is straightforward to find general expressions of renormalization group equations
\begin{eqnarray}
\frac{\mu}{e^2} \frac{de^2}{d\mu}
&=& -\epsilon + \frac{\mu}{Z_{a2}} \frac{dZ_{a2}}{d\mu} - \frac{2-\epsilon}{2} \frac{\mu}{Z_{b2}/Z_{b1}} \frac{d(Z_{b2}/Z_{b1})}{d\mu}, \nn
\frac{\mu}{\lambda} \frac{d\lambda}{d\mu}
&=& -\epsilon - \frac{\mu}{Z_\lambda} \frac{dZ_\lambda}{d\mu} + 2\frac{\mu}{Z_{b2}} \frac{dZ_{b2}}{d\mu} \nn
&& -\frac{2-\epsilon}{2} \frac{\mu}{Z_{b2}/Z_{b1}} \frac{d(Z_{b2}/Z_{b1})}{d\mu}, \nn
\frac{\mu}{\zeta_a^2} \frac{d\zeta_a^2}{d\mu}
&=& - \frac{\mu}{Z_{b2}/Z_{b1}} \frac{d(Z_{b2}/Z_{b1})}{d\mu} + \frac{\mu}{Z_{a2}/Z_{a1}} \frac{d(Z_{a2}/Z_{a1})}{d\mu}. \nn
\end{eqnarray}
Introducing counter terms into renormalization factors in the above equations, we obtain $\beta-$functions in $d=2$
\begin{eqnarray}
\beta_e
&\equiv& \mu\frac{de^2}{d\mu}
= e^2 \left( -1 + \frac{e^2}{96\pi^2} + \frac{e^2}{16\pi^2N} \frac{\log \zeta_a^2}{\zeta_a^2 -1} \right) \nn
\beta_\lambda
&\equiv& \mu\frac{d\lambda}{d\mu}
= \lambda \bigg(-1 + \frac{(N+5)\lambda}{16\pi^2N} + \frac{e^4}{16\pi^2 N \lambda \zeta_a^2} \nn &&  - \frac{3}{16\pi^2 N} \frac{\log \zeta_a^2}{\zeta_a^2 -1} e^2\lambda \bigg) \nn
\beta_{\zeta_a}
&\equiv& \mu \frac{d\zeta_a^2}{d\mu}
= \zeta_a^2 \left( \frac{e^2}{8\pi^2N} \frac{\log \zeta_a^2}{\zeta_a^2 -1} + \frac{e^2}{96\pi^2} - \frac{e^2}{96\pi^2  \zeta_a^2} \right), \nn
\label{RG_Bosonization}
\end{eqnarray}
which show how renormalized parameters flow as a function of the scaling parameter $\mu$. Comparing these renormalization group equations with those of both holon and gauge-field parts in Eq. (\ref{RG_Equations_Conformal_Symmetry}), we find that they are essentially identical in the physical point of view. See Fig. \ref{fig:FP_bosonization}. We note that the coupling constant $\lambda$ has a stable fixed point in the case of $N>N_c$. Here, the critical holon flavor number is $N_c=111$. All fixed point values are summarized in Table \ref{table:FPs2}.
\begin{table}[b]
\caption{
Fixed point values.
}
\begin{ruledtabular}
\begin{tabular}{cccc}
 & $N=1$ & $N=150$ & $N\rightarrow \infty$ \\
\colrule
$e_*^2/(8\pi^2)$     & 0.4841  & 11.52    & 12 \\
$\lambda_*/(8\pi^2)$ & -       & 1.581    & 2 \\
$\zeta_a^2$          & 0.02029 & 0.9232   & 1 \\
\end{tabular}
\end{ruledtabular}
\label{table:FPs2}
\end{table}
%
%

\section{Justification of the Luttinger-liquid physics in the critical spinon dynamics: SU(2) gauge theory point of view}

As discussed in section III-A, we took into account only the particle-hole channel for the decomposition of the Hubbard interaction, referred to as the U(1) slave-rotor theory. If the particle-particle channel is also introduced, we obtain an effective theory, referred to as the SU(2) slave-rotor representation \cite{KSKim_SU2SR}. Using the Nambu-spinor representation $\psi_{i}= \big( c_{i\uparrow}, c_{i\downarrow}^\dagger \big)^T$ and performing the Hubbard-Stratonovich transformation for both the particle-particle and particle-hole channels, we arrive at the following action:
\begin{eqnarray}
S
&=& \int_0^\beta\!d\tau \bigg[ \sum_i \psi_i^\dagger (\partial_\tau - \mu \tau_z - i\vec \Omega_i \cdot \vec \tau) \psi_i \nn
&& - t \sum_{ij} ( \psi_i^\dagger \tau_z \psi_j + H.c. ) + \frac{3}{4U} \sum_i \tr (\vec \Omega_i \cdot \vec \tau)^2 \bigg], \label{EFT_SU2}
\end{eqnarray}
where $\vec \tau = (\tau_x, \tau_y, \tau_z)$ are Pauli matrices, and $\vec \Omega_i$ is a Hubbard-Stratonovich field.

Similar to the U(1) slave-rotor theory, we decompose the Nambu-spinor field $\psi_i$ as
\begin{eqnarray}
\psi_i = Z_i^\dagger F_i,
\end{eqnarray}
where
\begin{eqnarray}
F_i = \begin{pmatrix} f_{i\uparrow} \\ f_{i\downarrow}^\dagger \end{pmatrix}, \ \ \
Z_i = \begin{pmatrix} z_{i\uparrow} & -z_{i\downarrow}^\dagger \\ z_{i\downarrow} & z_{i\uparrow}^\dagger \end{pmatrix} \in SU(2).
\end{eqnarray}
$f_{i\sigma}$ is a fermion field and $z_{i\sigma}$ is a bosonic field with the unimodular constraint of $|z_{i\uparrow}|^2 + |z_{i\downarrow}|^2 = 1$. Introducing this projective representation into Eq. (\ref{EFT_SU2}), redefining the Hubbard-Stratonovich field $\vec \Omega_i$ as $Z_i (\vec \Omega_i \cdot \vec \tau) Z_i^\dagger \rightarrow \vec \Omega_i \cdot \vec \tau$, and then shifting it as $\vec \Omega_i \cdot \vec \tau \rightarrow \vec \Omega_i \cdot \vec \tau - i Z_i \partial_\tau Z_i^\dagger$, we reach the following effective action
\begin{eqnarray}
S
&=& \int_0^\beta\!d\tau \bigg[ \sum_i F_i^\dagger \left( \partial_\tau - \mu Z_i \tau_z Z_i^\dagger - i \Omega_i^a \tau_a \right) F_i \nn
&& - t \sum_{ij} \left( F_i^\dagger Z_i \tau_z Z_j^\dagger F_j + H.c. \right) \nn
&&+ \frac{3}{4U} \sum_i \tr \left( \Omega_i^a \tau_a - iZ_i \partial_\tau Z_i^\dagger \right)^2 \nn
&&+ \sum_i \lambda_i \tr \left( Z_i^\dagger Z_i - 1 \right)\bigg],
\label{eq:SU2_action1}
\end{eqnarray}
where $a=x,y,z$ with the Einstein convention for the summation of $a$. The last term comes from the rotor constraint of $|z_{i\uparrow}|^2 + |z_{i\downarrow}|^2 = 1$, where $\lambda_i$ is a Lagrange multiplier field. For details in the derivation, we would like to refer to Ref. \cite{KSKim_SU2SR}.

Similar to the U(1) case, we consider the following mean-field ansatz:
\begin{eqnarray}
&&\big< Z_i \tau_z Z_j^\dagger \big> = X \tau_z e^{-iA_{ij}^a \tau_a}, ~~~  \big< F_i F^\dagger_j \big> = -Y \tau_z e^{-i A_{ij}^a \tau_a} , \nn
\end{eqnarray}
where $X$ and $Y$ are amplitudes of hopping parameters, determined by self-consistent equations of order parameters, and $e^{-iA_{ij}^a\tau_a}$'s are SU(2) gauge fluctuations in the lattice construction, taken into account beyond the mean-field approximation. Substituting this ansatz into the action Eq. (\ref{eq:SU2_action1}), we obtain
\begin{eqnarray}
S
&=& \int_0^\beta\!d\tau \bigg[ \sum_i F_i^\dagger \left( \partial_\tau - \tilde \mu \tau_z - i \Omega_i^a \tau_a \right) F_i \nn
&& - tX \sum_{ij} \left( F_i^\dagger \tau_z e^{-iA_{ij}^a \tau_a} F_j + H.c. \right) \nn
&&+ \frac{3}{4U} \sum_i \tr \left( \Omega_i^a \tau_a - iZ_i \partial_\tau Z_i^\dagger \right)^2 + \tilde\lambda \sum_i \tr Z_i^\dagger Z_i \nn
&&- tY \sum_{ij} \tr \left( Z_i^\dagger \tau_z e^{-iA_{ij}^a \tau_a} Z_j \tau_z + H.c. \right) \bigg]\nn
&&- \lambda \beta N - 2z XY \beta N ,
\label{eq:SU2_action2}
\end{eqnarray}
where $\lambda_{i}$ is chosen to be uniform. $z$ in the last line is the coordination number. $\tilde \mu$ and $\tilde \lambda$ are modified values for the chemical potential of spinons and the mass of holons, respectively, due to the on-site interaction $-\mu \sum_i F_i^\dagger Z_i \tau_z Z_i^\dagger F_i$ in the first line of Eq. (\ref{eq:SU2_action1}).

Taking into account the continuum limit, we reach the following expression for the SU(2) slave-rotor theory in the spin-liquid ansatz
\begin{eqnarray}
S
&=& \int_0^\beta\!d\tau\int\!d^2x \bigg[ F^\dagger \left( \partial_\tau - \tilde \mu \tau_z - ig\Omega^a \tau_a \right) F \nn
&&- tX F^\dagger  \tau_z \big( \vec \nabla - ig\vec A^a \tau_a \big)^2 F  \nn
&&- \frac{3}{4U} \tr \left( Z^\dagger \partial_\tau Z - ig\Omega^a \tau_a \right)^2 + \tilde\lambda \ \tr Z^\dagger Z \nn
&&- tY \tr \left\{ Z^\dagger \tau_z \big( \vec\nabla - ig\vec A^a\tau_a \big)^2 Z \tau_z \right\} \nn
&&- \frac{1}{4} F_{\mu\nu}^a F_{\mu\nu}^a \bigg].
\end{eqnarray}
Here, we dropped last two terms in Eq. (\ref{eq:SU2_action2}), assuming that the mean-field values of $\lambda$, $X$, and $Y$ are determined to give constant contributions. In addition, we include the kinetic term of gauge-field fluctuations, expected to result from high-energy fluctuations of spinons and holons, where $A_\mu^a$ ($\mu=0,1,2$) with $A^a_0 = \Omega^a$ and rescaling of $A\rightarrow gA$. $F_{\mu\nu}^a = \partial_\mu A_\nu^a - \partial_\nu A_\mu^a + gf^{abc} A_\mu^b A_\nu^c$ is a field strength with an interaction coupling constant $g$. $f^{abc}$ is a structure constant defined by commutators of generators $[T_a, T_b] = if^{abc} T_c$, where $T_a = \tau_a/2$ in our case, and thus $f^{abc}$ is given by the Levi-Civita symbol $\epsilon^{abc}$. The spatial dimension is two and the arrowed vector is a two-dimensional vector.

In order to quantize the non-Abelian gauge theory, we take into account both terms involved with gauge fixing and ghost fields \cite{Callan_Symanzik_Eqs}:
\begin{eqnarray}
S_{add} = -\frac{1}{2\xi} \int\! d^3x \big( \partial_i A_i^a \big)^2 - \int\! d^3x \ \bar c^a \partial_i D_i^{ac} c^c,
\end{eqnarray}
where $\xi$ is a parameter that fixes the gauge field propagator, $c^a$ are fermion ghost fields, and $D_i^{ac} = \delta^{ac} \partial_i + gf^{abc}A^b_i$.

Following the U(1) slave-rotor theory, we write down an effective field theory in a double-patch construction. Expanding the dispersion near the Fermi surface, we obtain the free-part of the spinon dynamics as follows
\begin{eqnarray}
S
= \int\!\frac{d^3k}{(2\pi)^3} F_s^\dagger(k) \big[ ik_0 + (sv_F k_1 + tX k_2^2)\tau_z \big] F_s(k),
\end{eqnarray}
where $s=\pm$ is the patch index, $v_F = 2tXK_F$, and $K_F$ is Fermi wave vector. In order to construct $(1+1)D-$type Dirac representation, we combine two patches as
\begin{eqnarray}
\Psi(k) = \begin{pmatrix} F_+(k) \\ \left( F_-^\dagger \right)^T (-k) \end{pmatrix},
\end{eqnarray}
where $T$ is the transpose operator. $\left( F_-^\dagger \right)^T(-k)$ means a column vector with the components of $f_-^\dagger(-k)$ and $f_-(k)$. Resorting to this non-minimal representation, we rewrite the above action in the (1+1)-dimensional Dirac form:
\begin{eqnarray}
S
= \int\!\frac{d^3k}{(2\pi)^3} \ \bar\Psi(k) \big( i\gamma_0 k_0 + i\gamma_1 \delta_k \big) \Psi(k) ,
\end{eqnarray}
where
\begin{eqnarray}
\gamma_0 =
	\begin{pmatrix}
	0    & -iI_2 \\
	iI_2 & 0
	\end{pmatrix}, \ \ \
\gamma_1 =
	\begin{pmatrix}
	0      & \tau_z\\
	\tau_z & 0
	\end{pmatrix}
\end{eqnarray}
with a two by two identity matrix $I_2$, $\delta_k = v_F k_1 + tX k_2^2$.

The interaction term of the spinon-gauge field in the patch description is
\begin{eqnarray}
S_{FA}
= -v_F g \int_{k,q} s F^\dagger_s(k+q) \tau_z A^a(q) \tau_a F_s(k),
\end{eqnarray}
where we imposed the Coulomb gauge condition on $\vec A^a$ so that $\vec q \cdot \vec A^a(q)=0$. Then, $\vec A^a(q)$ has only one component in two dimensions, denoted by $A^a(q)$. Resorting to the spinor representation of $\Psi$, the interaction term can be written as
\begin{eqnarray}
S_{FA}
&=& iv_F e \int_{k,q} \bar\Psi(k+q) \gamma_0\gamma_1 \Big( A^x(q) \tau_x \bar I_4 \nn &+& A^y(q)\tau_y I_4 + A^z(q) \tau_z I_4 \Big) \Psi(k),
\end{eqnarray}
where $I_4$ is a four by four identity matrix and $\bar I_4 = \mathrm{diag}(I_2, -I_2)$.

Now, it is straightforward to consider the dimensional regularization. Extending the co-dimension of the spinon Fermi surface, and perform the tree-level scaling analysis, we obtain the following scaling transformation:
\begin{eqnarray}
&&{\bm K} = \frac{{\bm K}'}{s}, \
k_{d-1} = \frac{k_{d-1}'}{s}, \
k_d = \frac{k_d}{\sqrt{s}}, \nn
&&\Psi(k) = s^{\Delta_\Psi} \Psi'(k'), \ \ \
\Delta_\Psi = \frac{d}{2} + \frac{3}{4}, \nn
&&A^a(q) = s^{\Delta_A} {A^a}'(q'), \ \ \
\Delta_A = \frac{d}{2} + \frac{3}{4},\nn
&&c(k) = s^{\Delta_c}c'(k'), \ \ \
\Delta_c = \frac{d}{2} + \frac{3}{4}, \nn
&&g=s^{\Delta_g} g', \ \ \
\Delta_g = \frac{d}{2} - \frac{5}{4},
\end{eqnarray}
essentially the same as the U(1) case. Here, the scaling transformation of the coupling $g$ is deduced from the spinon-gauge field coupling term.
However, the difference of SU(2) theory to U(1) theory is that the gauge fields are interacting with themselves given by the same coupling constant $g$. In order to have a consistent description, the scaling transformation obtained from the self-interaction term of the gauge field should be the same to that in the above equation, even in the tree level. From both the $A^3-$ and $A^4-$ interaction terms, we have
\begin{eqnarray}
&&g=s^{\Delta_{g3}} g', \ \ \
\Delta_{g3} = \frac{d}{2} - \frac{1}{4}, \nn
&&g=s^{\Delta_{g4}} g', \ \ \
\Delta_{g4} = \frac{d}{2} - \frac{3}{4} .
\end{eqnarray}
As a result, $\Delta_{g3}$, $\Delta_{g4}$, and $\Delta_g$ are all different to each other in any dimensions. We also find
\begin{eqnarray}
g=s^{\Delta_{cA}}g', \ \ \
\Delta_{cA} = \frac{d}{2}- \frac{1}{4}
\end{eqnarray}
in the ghost-gauge field coupling term, different from $\Delta_g$. These observations lead us to conclude that the scaling transformation to preserve the spinon Fermi surface cannot be consistent with the SU(2) gauge symmetry. On the other hand, releasing the stability condition for the spinon Fermi surface, the critical spinon dynamics is described by the one-dimensional Luttinger-liquid physics. The critical holon dynamics is still described by the two-dimensional band description. An important point is that the Lorentz invariance is fully respected when we preserve the holon dispersion relation across the metal-insulator transition. As a result, the inconsistency between the scaling transformation and the SU(2) gauge symmetry does not occur. The SU(2) gauge symmetry seems to suggest one-dimensional Luttinger-liquid physics for critical spinon dynamics at UV although this question should be addressed more carefully near future.

\section{Summary and Discussion}

\subsection{Summary}

In this study we investigated how a spinon Fermi surface becomes destabilized to result in the emergence of one-dimensional spin dynamics, based on the perturbative theoretical framework. Actually, we could obtain such a nonperturbative phenomenon based on the renormalization group analysis within the scheme of graphenization of the Fermi-surface problem. An essential point is that the spinon Fermi surface becomes flattened along the direction of the Fermi surface already at the tree level near the spin-liquid Mott quantum criticality. As a result, quantum critical dynamics of spinons is described by one-dimensional relativistic spectrum at UV, i.e., the physics of Luttinger liquid. Then, gapless low lying spin-singlet fluctuations described by U(1) gauge fields cannot be damped due to the presence of pseudogap physics. Interaction effects are much enhanced to cause Luttinger-liquid physics to the spinon dynamics at the spin-liquid Mott quantum critical point of IR. On the other hand, critical charge fluctuations are governed by an IXY fixed point above a critical value of the holon flavor number.

We believe that essential ingredients in the effective field theory are extra, or more precisely, actual critical degrees of freedom in addition to ``generic scale invariance". Here, we use the term of generic scale invariance in the sense of that used in Ref. \cite{Belitz_GSI_RMP}. The generic scale invariance is a feature of the U(1) spin-liquid state, where such a phase is identified with an interacting stable fixed point in the renormalization group analysis. Additional or actual critical degrees of freedom are given by holon excitations, physically speaking, fluctuations of zero-sound modes, describing a metal-insulator Mott transition from a Fermi-liquid phase to a spin-liquid state. Here, we ask possible UV fixed points that we should start from for the renormalization group analysis. One may choose the U(1) spin-liquid critical fixed point with a stable spinon Fermi surface as a starting UV fixed point. However, we are suggesting another possibility here in order to discuss the metal-insulator transition: If we go to the Mott critical point from the Fermi-liquid phase, it may be better to perform the scaling analysis which fits to the critical holon sector. In other words, we start from a different UV fixed point. As a result, the Lorentz invariance emerges at the tree level, which does not allow Landau damping. The absence of Landau damping is responsible for the destabilization of a spinon Fermi surface.

Now, the question is how generic this feature is. Suppose two dimensional interacting electrons in the presence of nonmagnetic disorders. Then, this system flows into a diffusive Fermi-liquid fixed point, where the generic scale invariance occurs \cite{Belitz_GSI_RMP}. Reducing the density of electrons, experiments tell us that ferromagnetic spin fluctuations seem to appear in the vicinity of a metal-insulator transition. In order to describe these ferromagnetic spin fluctuations, one may introduce an additional order parameter as a critical theory. Here, a question arises: how should we take the scaling analysis in the tree level? Which fixed points should we resort to for the scaling analysis: the diffusive fixed point or such a magnetic ``quantum critical" (more precisely, instability) point? We can consider another situation. We revisit the U(1) spin-liquid state. Now, we take into account a spin-density-wave instability inside the spin-liquid phase for a generic spinon Fermi surface. Here, the spin-density-wave transition can occur in the $2 \bf{k}_{F}$ momentum channel, where $\bf{k}_{F}$ is a spinon Fermi momentum. Which fixed points should we start from: the U(1) spin-liquid phase or the spin-density-wave quantum critical point? We suspect that spinons may become localized in the second case while antiferromagnetic critical spin fluctuations can be itinerant. We expect that this physical situation would realize a two-fluid model description. This speculation should be investigated more sincerely later.

\subsection{Physical picture: A scenario for a renormalization group flow from the U(1) spin-liquid fixed point with a stable spinon Fermi surface to the emergent Luttinger-liquid physics of spinons}

An essential question in the present study is how the spinon Fermi surface disappears, approaching the spin-liquid to Fermi-liquid Mott quantum critical point from the U(1) spin-liquid state with the spinon Fermi surface. Here, we did not find the renormalization group flow from the U(1) spin-liquid fixed point of S.-S. Lee \cite{SSL_Dimensional_Regularization_Nematic} to the spin-liquid Mott quantum critical point of ours within the scheme of graphenized dimensional regularization. In order to verify this renormalization group flow, we suggest an idea to perform the renormalization group analysis with the curvature term, assuming the relativistic scaling transformation which leads the holon dynamics to be invariant. Although the curvature term is irrelevant in the tree-level scaling analysis, there may appear quantum corrections of ``anti-screening" to cause a run-away flow. In other words, we speculate that there is a critical value of the curvature term or effective band mass: When the effective mass is more than a critical value, the velocity is renormalized to vanish, i.e., showing Mott localization along the transverse direction of the Fermi surface near the quantum critical point. On the other hand, when the effective mass is less than the critical value, the spinon Fermi surface would be stabilized. This physical picture may be possible if the Landau damping term in U(1) gauge fluctuations is controlled by the evolution of the curvature term.

%
%

\subsection{A possible connection between dynamical mean-field theory and U(1) spin-liquid theory}

Dynamical mean-field theory assumes the emergence of localized magnetic moments in the vicinity of a metal-insulator transition \cite{DMFT_Review}. An insulating state within this description is given by self-consistently generated localized magnetic moments decoupled from itinerant electrons. A metallic phase is described by screening of such preexisting localized magnetic moments, nothing but self-consistently describing pseudogap-like Kondo effect \cite{Discussion_Vlad}. Such emergent localized magnetic moments carry extensive entropy. Thus, they play the role of the source of strong inelastic scattering in dynamics of itinerant electrons, responsible for non-Fermi liquid physics near the local-moment Mott quantum criticality. Actually, the dynamical mean-field theory could explain the mirror-shaped quantum critical scaling behavior for electrical resistivity in the vicinity of various metal-insulator transitions quite surprisingly \cite{Vlad_Recent_Works_MQCP}. Unfortunately, such a theoretical framework gives an unsatisfactory description on how emergent localized magnetic moments are screened to reduce huge entropy at low temperatures in the insulating phase. On the other hand, characteristic features of the insulating phase in $\kappa-$class organic salts turn out to be well described by spin-liquid physics, more precisely, the U(1) spin-liquid state with a spinon Fermi surface \cite{Kappa_BEDT_Review}. It would be interesting to investigate whether or not the emergence of one-dimensional spinon dynamics at the spin-liquid Mott quantum critical point serves a meaningful connection from spin-liquid theory to dynamical mean-field theory for Mott quantum criticality.

\section*{Acknowledgement}

This study was supported by the Ministry of Education, Science, and Technology (No. NRF-2015R1C1A1A01051629 and No. 2011-0030046) of the National Research Foundation of Korea (NRF) and by TJ Park Science Fellowship of the POSCO TJ Park Foundation. This work was also supported by the POSTECH Basic Science Research Institute Grant (2016). We would like to appreciate fruitful discussions in the APCTP Focus program ``Lecture series on Beyond Landau Fermi liquid and BCS superconductivity near quantum criticality" in 2016. K.S.K appreciates fruitful discussions with R. Narayanan on how to construct a DMFT theory for spin liquid physics at low temperatures.

\appendix

\section{Polarization function $\Pi_1$, Fermion self-energy $\Sigma^f$, and Boson self-energy $\Sigma^b$ in renormalization group analysis I}
\label{App:polarization1}

In this appendix, we evaluate the self-energies and polarizations needed in RG analysis of Sec. \ref{Sec:RG_I}.

\begin{figure}[h]
\includegraphics[width=7cm]{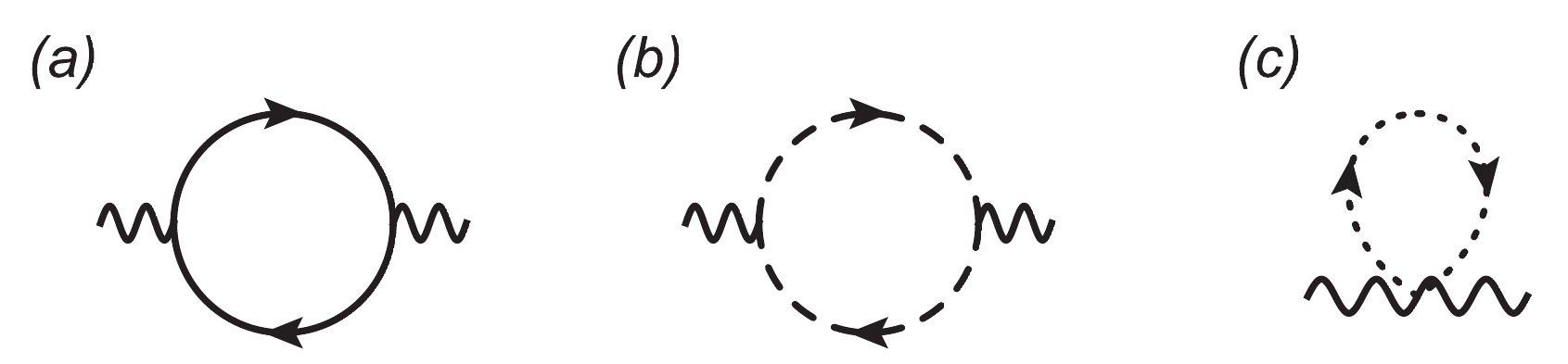}
\caption{
Polarization diagrams from spinons (a) and holons (b) $\&$ (c).
}
\label{fig:App_pol}
\end{figure}

There are three diagrams for the polarization in the one-loop level, but the fermion bubble is the order of $\mathcal{O}(N^0)$ while two Boson bubbles are the order of $\mathcal{O}(1/N)$. In this respect we consider only the fermion bubble in the gauge field propagator.

The spinon bubble diagram (Fig. \ref{fig:App_pol} (a)) is
\begin{eqnarray}
\Pi_f(q)
&=& - \left( \frac{ie\mu^{\epsilon\over2}}{\sqrt{N}} \right)^2 \int\!\frac{d^{d+1}k}{(2\pi)^{d+1}} \nn
&&\times \tr \Big[ G_0^\psi(k) \gamma_5 \gamma_{d-1} G_0^\psi(k+q) \gamma_5 \gamma_{d-1} \Big] \nn
&&= -\frac{e^2\mu^\epsilon}{N} \beta_d \frac{|{\bm Q}|^{d-1}}{|q_d|},
\end{eqnarray}
where
\begin{eqnarray}
\beta_d = \frac{\Gamma^2\left( {d\over2} \right)}{2^d \pi^{d-1\over2} \left| \cos {\pi d\over2} \right| \Gamma\left( {d-1\over2} \right) \Gamma(d) }.
\end{eqnarray}
The minus sign in the first line is due to the fermion loop.


\begin{figure}[h]
\includegraphics[width=7cm]{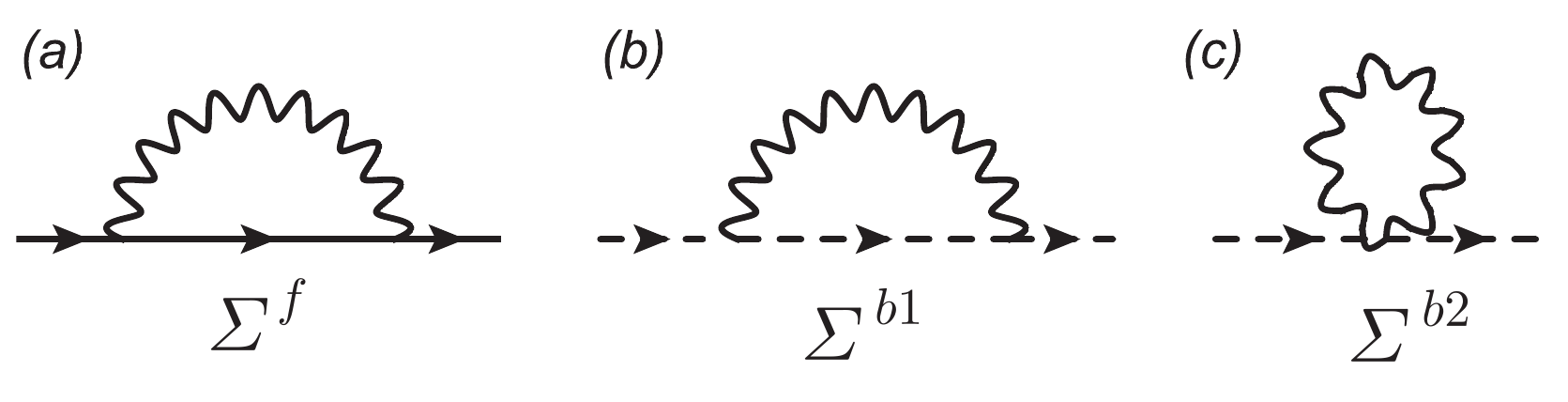}
\caption{
Self-energy corrections for spinons (a) and holons (b) $\&$ (c).
}
\label{fig:App_SE}
\end{figure}

The spinon self-energy (Fig. \ref{fig:App_SE} (a)) is given by
\begin{eqnarray}
\Sigma^f(k)
&=& \left( \frac{ie\mu^{\epsilon\over2}}{\sqrt{N}} \right)^2 \int\!\frac{d^{d+1}q}{(2\pi)^{d+1}} \nn
&&\times G_0^a(q) \gamma_5 \gamma_{d-1} G_0^\psi(k+q) \gamma_5 \gamma_{d-1} \nn
&=& - \frac{ie^{4\over3}\mu^{2\epsilon\over3}}{N^{2\over3}} \frac{\beta_\Sigma^f}{\beta_d^{1\over3}} |{\bm K}|^{2d-5\over6} ({\bm \Gamma}\cdot {\bm K}),
\end{eqnarray}
where
\begin{eqnarray}
\beta_\Sigma^f
= \frac{\Gamma\left( {5-2d\over6} \right) \Gamma\left( {d-1\over3} \right) \Gamma\left( {d\over2} \right)}{3\sqrt{3} \ 2^{d-1} \pi^{d\over2} \Gamma\left( {d-1\over6} \right) \Gamma\left( {5d-2\over6} \right)}.
\end{eqnarray}
This gives
\begin{eqnarray}
\Sigma^f(k)
&=& - \frac{e^{4\over3}}{N^{2\over3}}\frac{u_1}{\epsilon} (i{\bm \Gamma} \cdot {\bm K}),
\end{eqnarray}
where
\begin{eqnarray}
u_1 = \frac{1}{2^{3\over2} 3^{3\over2} \pi^{3\over4} \Gamma\left( {3\over4} \right) \beta_{5\over2}^{1\over3}}.
\end{eqnarray}

The holon self-energy (Fig. \ref{fig:App_SE} (b)) is
\begin{eqnarray}
\Sigma^{b1}(k)
&=& -\left( -\frac{2e\mu^{\epsilon\over2}}{\sqrt{N}} \right)^2 \int\!\frac{d^{d+1}q}{(2\pi)^{d+1}} \nn
&&\times (\vec k \cdot \hat t_q)^2 G_0^a(q) G_0^b(k+q).
\end{eqnarray}
Here, we show that this term does not give the $1/\epsilon$-divergence. Since the inner-product term in the integrand does not affect the $1/\epsilon$-divergence, we consider the following integral:
\begin{eqnarray}
I
&\equiv& \int\!\frac{d^{d+1}q}{(2\pi)^{d+1}} \ \frac{1}{q_d^2 + e^2\mu^\epsilon \beta_d \frac{|{\bm Q}|^{d-1}}{|q_d|}} \nn
&&\hspace{70pt} \times \frac{1}{\zeta_b ({\bm Q} + {\bm K})^2 + (\vec q + \vec k)^2} \nn
&=& \frac{1}{2} \int\!\frac{d{\bm Q} dq_d}{(2\pi)^{d}} \ \frac{1}{q_d^2 + e^2\mu^\epsilon \beta_d \frac{|{\bm Q}|^{d-1}}{|q_d|}} \nn
&&\hspace{40pt} \times \frac{1}{\Big[ \zeta_b ({\bm Q} + {\bm K})^2 + (q_d + k_d)^2 \Big]^{1\over2}}.
\end{eqnarray}
If $q_d^2$-term is dominant than Landau damping in the gauge-field propagator, we have no $1/\epsilon$ divergence. On the other hand, if Landau damping is dominant, $q_d$ is cut-offed by $|{\bm Q}|^{d-1\over3}$. Considering the approximation of the gauge propagator proportional to $|q_d|/|{\bm Q}|^{d-1}$ and integrating over $q_d$, we obtain
\begin{eqnarray}
I
&\sim& \int\!dQ \ Q^{d-2} \frac{\sqrt{Q^{2} + Q^{2d-2\over3}}}{Q^{d-1}} \nn
&=& \int\!dQ \ \sqrt{1+Q^{2d-8\over3}}.
\end{eqnarray}
The first term is not regularized by the dimension $d$, allowing us to neglect it. The second term can be regularized by the dimension but near $d=7$. So, the integral does not have the $1/\epsilon$-divergence near $d={5\over2}$.


The second boson self-energy term (Fig. \ref{fig:App_SE} (c)) is given by
\begin{eqnarray}
\Sigma^{b2}(k)
&=& \frac{e^2\mu^\epsilon}{N} \int\!\frac{d^{d+1}q}{(2\pi)^{d+1}} \frac{1}{q_d^2 + e^2\mu^\epsilon\beta_d \frac{|{\bm Q}|^{d-1}}{|q_d|}} \nn
&=& \frac{2}{3\sqrt{3}} \frac{e^2\mu^\epsilon}{N} \Lambda_{d-1} \int\!\frac{d^{d-1}Q}{(2\pi)^{d-1}} \frac{1}{\left({\bm Q}^2\right)^{d-1\over6}} \nn
&=& 0,
\end{eqnarray}
which vanishes due to the Veltman's formula \cite{Veltman_Formula}. $\Lambda_{d-1}$ is a momentum cutoff in the $q_{d-1}$ direction. There is no $1/\epsilon$-divergence.

\section{Derivation of Callan-Symanzik equation}
\label{App:CS_Eq}

In this appendix, we derive Callan-Symanzik equation for the model in Sec. \ref{Sec:RG_II}. A bare $(m+n+l)$-point Green's function is defined as
\begin{eqnarray}
&&\Big< \bar\psi_B(k_{B,1}) \cdots \bar\psi_B(k_{B,m}) \psi_B(k_{B,m+1}) \cdots \psi_B(k_{B,2m}) \nn
&&\times b^\dagger_B(k_{B,2m+1}) \cdots b^\dagger_B(k_{B,2m+n}) b_B(k_{B,2m+n+1}) \nn && \cdots b_B(k_{B,2m+2n}) a_B(k_{B,2m+2n+1}) \cdots a_B(k_{B,2m+2n+2l}) \Big> \nn
&&= G^{(m,n,l)}_B \Big( \{ k_{B,i} \}; e_B, \lambda_B, \zeta_{\psi, B}, \zeta_{a, B} \Big) \ \delta^{(d+1)} \Big(\{ k_{B,i} \}\Big) , \nn
\end{eqnarray}
and a renormalized Green's function is
\begin{eqnarray}
&&\Big< \bar\psi(k_1) \cdots \bar\psi(k_m) \psi(k_{m+1}) \cdots \psi(k_{2m}) \nn
&&\times b^\dagger(k_{2m+1}) \cdots b^\dagger(k_{2m+n}) b(k_{2m+n+1}) \cdots b(k_{2m+2n}) \nn
&&\ \times a(k_{2m+2n+1}) \cdots a(k_{2m+2n+2l}) \Big> \nn
&&= G^{(m,n,l)} \Big( \{ k_i \}; e, \lambda, \zeta_\psi, \zeta_a, \mu \Big) \ \delta^{(d+1)} \Big( \{ k_i \} \Big) ,
\end{eqnarray}
where the relation between bare and renormalized Green's functions is
\begin{eqnarray}
&&G^{(m,n,l)} \Big( \{ k_i \}; e, \lambda, \zeta_\psi, \zeta_a, \mu \Big) \nn
&&\hspace{30pt} = Z_{\psi}^{-m} Z_{b}^{-n} Z_{a}^{-n} \left(\frac{Z_{b2}}{Z_{b1}}\right)^{d-1\over2} \nn
&&\hspace{30pt} \times G^{(m,n,l)}_B \Big( \{ k_{B,i} \}; e_B, \lambda_B, \zeta_{\psi, B}, \zeta_{a, B} \Big).
\end{eqnarray}

The bare correlation function should not depend on the energy scale $\mu$, given by $\mu{d\over d\mu}G_B = 0$. Then, we obtain the following differential equation for the renormalized correlation function
\begin{eqnarray}
&& \bigg[ \mu \frac{\partial}{\partial \mu} + (1-z) {\bm K} \cdot \nabla_{\bm K} \nn && + \beta_e \frac{\partial}{\partial e^2} + \beta_\lambda \frac{\partial}{\partial \lambda} + \beta_{\zeta_\psi} \frac{\partial}{\partial \zeta_\psi} + \beta_{\zeta_a} \frac{\partial}{\partial \zeta_a} \nn && + 2m\eta_\psi + 2n\eta_b + 2l\eta_a - (d-1)(1-z) \bigg] \nn && G^{(m,n,l)} \Big( \{ k_i \}; e, \lambda, \zeta_\psi, \zeta_a, \mu \Big) = 0, \label{CSE_I}
\end{eqnarray}
where the $\beta$-functions $\beta_g$, $g=e, \lambda, \zeta_\psi, \zeta_a$ are defined as $\beta_g = d g / d \log \mu$. The dynamical critical exponent $z$, and anomalous scaling dimensions $\eta_i$, $i=\psi, b, a$ are given by
\bqa
&&z = 1- \frac{1}{2} \frac{\mu}{Z_{b2}/Z_{b1}} \frac{\partial Z_{b2}/Z_{b1}}{\partial \mu}, \nn
&&\eta_\psi = \frac{1}{2} \frac{\mu}{Z_\psi} \frac{\partial Z_\psi}{\partial \mu}, \ \ \
\eta_b = \frac{1}{2} \frac{\mu}{Z_b} \frac{\partial Z_b}{\partial \mu}, \ \ \
\eta_a = \frac{1}{2} \frac{\mu}{Z_a} \frac{\partial Z_a}{\partial \mu}. \nn
\eqa

From the definition of $(m+n+l)$-point Green's function, the engineering scaling dimension of $G^{(m,n,l)}$ given by
\begin{eqnarray}
&& G(sk;\mu) = s^{D} G(k; \mu/s)
\end{eqnarray}
is
\begin{eqnarray}
&&D = -2m \frac{d+2}{2} - 2n \frac{d+3}{2} - 2l \frac{d+3}{2} + (d+1) . \nn
\end{eqnarray}
As a result, we obtain
\begin{eqnarray}
%
%
&&\left( {\bm K}_i \cdot \nabla_{\bm K} + \vec k \cdot \nabla_{\vec k} + \mu \frac{\partial}{\partial \mu} - D \right) G = 0.
\end{eqnarray}

Combining this equation with the previous equation (\ref{CSE_I}), we reach the following expression of a differential equation for a $(m+n+l)$-point correlation function, which shows the evolution as a function of the energy scale,
\begin{eqnarray}
&&\bigg[ z {\bm K}_i \cdot \nabla_{{\bm K}_i} + \vec k_i \cdot \nabla_{\vec k_i} \nn && - \beta_e \frac{\partial}{\partial e^2} - \beta_\lambda \frac{\partial}{\partial \lambda} - \beta_{\zeta_\psi} \frac{\partial}{\partial \zeta_\psi} - \beta_{\zeta_a} \frac{\partial}{\partial \zeta_a} \nn && - 2m \left( - \frac{5-\epsilon}{2} + \eta_\psi \right) - 2n \left( - \frac{6-\epsilon}{2} + \eta_b \right) \nn && - 2l \left( - \frac{6-\epsilon}{2} + \eta_a \right) - \{ z(2-\epsilon) + 2 \} \bigg] \nn && G^{(m,n,l)} \Big( \{ k_i \}; e, \lambda, \zeta_\psi, \zeta_a, \mu \Big) = 0.
\end{eqnarray}
Here, the dimensions is $d=3-\epsilon$. This is Callan-Symanzik equation for our model, Eq. (\ref{eq:CS_eq}) in the main text.



\begin{thebibliography}{9}
\bibitem{Patch_Construction} In the renormalization group analysis for the Landau's Fermi-liquid state, the transverse momentum is dimensionless, described by an angular coordinate. On the other hand, the transverse momentum in quantum critical metals (critical Fermi surfaces due to critical order parameter fluctuations) acquires an anomalous scaling dimension due to the presence of singular self-energy corrections, as shown in Ref. \cite{SSL_Breakdown_N_Before}. Suppose two nearby patches, described by two transverse momenta. Since the transverse momentum has an anomalous scaling dimension, it turns out that the area of the overlap region between two nearby patches decreases much faster than that of the patch region defined by a fermion-momentum cutoff. In other words, the anomalous scaling for the transverse direction leads each patch decoupled asymptotically at low energies in momentum space. This emergent locality in momentum space allows us to consider a double-patch construction as a minimal model near quantum criticality. See Ref. \cite{SSL_Breakdown_N_Before} for more details.
\end{thebibliography}
\end{document}